\documentclass[]{emulateapj}
\usepackage{natbib,aas_macros}
\usepackage{hyperref}
\usepackage{graphicx}	
\usepackage{xspace}
\usepackage{xcolor}     

\newcommand{\ho}{$H_{0}$\xspace}                  
\newcommand{\cchp}{\,CCHP}                        
\newcommand{\hounit}{\,km\,s$^{-1}$\,Mpc$^{-1}$}  
\newcommand{\ocen}{$\omega$\,Centauri\,}          
\newcommand{\gaia}{\emph{Gaia}\xspace}            
\newcommand{\hip}{\emph{Hipparcos\xspace}}        
\newcommand{\hst}{\emph{HST}\xspace}              
\newcommand{\fgs}{\emph{HST}$+$\emph{FGS}\xspace}  
\newcommand{\wmap}{\emph{WMAP}\xspace}            
\newcommand{\sn}{SNe~Ia\xspace}                   

\shorttitle{The Carnegie-Chicago Hubble Program}
\shortauthors{Beaton et al.}
\begin{document}
\title{\textit{The Carnegie-Chicago Hubble Program}. I. AN INDEPENDENT APPROACH TO THE EXTRAGALACTIC DISTANCE SCALE USING ONLY POPULATION II DISTANCE INDICATORS\altaffilmark{*}}

\author{Rachael~L.~Beaton\altaffilmark{1,$\dagger$},
 Wendy~L.~Freedman\altaffilmark{2},
 Barry~F.~Madore\altaffilmark{1},
 Giuseppe~Bono\altaffilmark{3,4},
 Erika K.~Carlson\altaffilmark{1},
 Gisella Clementini\altaffilmark{5},
 Meredith J.~Durbin\altaffilmark{6},
 Alessia Garofalo\altaffilmark{5,7},
 Dylan Hatt\altaffilmark{2}, 
 In~Sung~Jang\altaffilmark{8},
 Juna~A.~Kollmeier\altaffilmark{1},
 Myung~Gyoon~Lee\altaffilmark{8},
 Andrew~J.~Monson\altaffilmark{9},
 Jeffrey~A.~Rich\altaffilmark{1},
 Victoria~Scowcroft\altaffilmark{1},
 Mark~Seibert\altaffilmark{1},
 Laura~Sturch\altaffilmark{1},
 and Soung-Chul~Yang\altaffilmark{10}}
 
 \altaffiltext{1}{The Observatories of the Carnegie Institution for Science, 813 Santa Barbara St., Pasadena, CA 91101, USA}
 \altaffiltext{2}{Department of Astronomy \& Astrophysics, University of Chicago, 5640 South Ellis Avenue, Chicago, IL 60637, USA}
 \altaffiltext{3}{Dipartimento di Fisica, Universit\'a di Roma Tor Vergata via Della Ricerca Scientifica 1, 00133, Roma, Italy}
 \altaffiltext{4}{INAF--Osservatorio Astronomico di Roma, Via Frascati 33, 00040 Monte Porzio Catone, Italy}
 \altaffiltext{5}{INAF-Osservatorio Astronomico di Bologna, via Ranzani 1, I-40127, Bologna, Italy}
 \altaffiltext{6}{Space Telescope Science Institute, 3700 San Martin Drive, Baltimore, MD 21218, USA}
 \altaffiltext{7}{Dipartimento di Fisica e Astronomia, Universit\'a di Bologna, Viale Berti Pichat 6/2, I-40127 Bologna, Italy}
 \altaffiltext{8}{Department of Physics \& Astronomy, Seoul National University, Gwanak-gu, Seoul 151-742, Korea}
 \altaffiltext{9}{Department of Astronomy \& Astrophysics, The Pennsylvania State University, 525 Davey Lab, University Park, PA 16802, USA}
 \altaffiltext{10}{Korea Astronomy and Space Science Institute (KASI), Daejeon 305-348, Korea}
 \altaffiltext{*}{Based on observations made with the NASA/ESA Hubble Space Telescope, obtained at the Space Telescope Science Institute, which is operated by the  Association of Universities for Research in Astronomy, Inc., under NASA contract NAS 5-26555. These observations are associated with programs \#13472 and \#13691.}
 \altaffiltext{$\dagger$}{E-mail: rbeaton@obs.carnegiescience.edu}

\begin{abstract}
We present an overview of the \emph{Carnegie-Chicago Hubble Program}, 
 an ongoing program to obtain a 3 per cent measurement 
 of the Hubble constant (\ho) using alternative methods to the traditional Cepheid distance scale.
We aim to establish a completely independent route to \ho using RR Lyrae variables, 
  the tip of the red giant branch (TRGB), and Type Ia supernovae (\sn). 
This alternative distance ladder can be applied to galaxies of any Hubble Type,
  of any inclination, and, utilizing old stars in low density environments, 
  is robust to the degenerate effects of metallicity and interstellar extinction.
Given the relatively small number of \sn host galaxies with independently measured distances,
  these properties provide a great systematic advantage in the measurement of \ho via 
  the distance ladder. 
Initially, the accuracy of our value of \ho will be set by the five 
  Galactic RR Lyrae calibrators with \emph{Hubble Space Telescope} Fine-Guidance Sensor parallaxes. 
With \gaia, both the RR Lyrae zero point and TRGB method 
  will be independently calibrated,
  the former with at least an order of magnitude more calibrators
  and the latter directly through parallax measurement of tip red giants.
As the first end-to-end ``distance ladder" completely independent of both Cepheid variables
 and the Large Magellanic Cloud, 
this path to \ho will allow for the high precision comparison at each rung of the traditional distance ladder
 that is necessary to understand tensions between this and other routes to \ho.
\end{abstract}
\keywords{distance scale -- cosmological parameters -- stars: variables: RR Lyrae -- stars: Population II}

\section{Introduction} \label{sec:intro} 

The determination of cosmological parameters has improved dramatically
 over the past two decades. 
The `factor-of-two' controversy over the value
 of the Hubble constant was resolved at the turn of the century \citep{freedman_2001}
 and only a decade later, quoted precisions are being claimed
 at the 3-5 per cent level \citep[e.g.,][]{riess_2011, freedman_2012, riess_2016}, 
 not only for \ho, but also for many of the other
 fundamental cosmological parameters.  
With measurements from \emph{Wilkinson Microwave Anisotropy Probe} (\wmap), 
 the \emph{Hubble Space Telescope} (\hst)~\ho Key Project, large scale surveys of \sn
 --- including the \emph{Supernovae, \ho, for the Equation of State of Dark Energy}
  \citep[$SH_{0}ES$; e.g.,][]{shoes,riess_2011,riess_2016} and the
  \emph{Carnegie Supernova Project} \citep[CSP; e.g.,][]{folatelli_2010,freedman_2009},  
 as well as from baryon acoustic oscillations made from the ground
  \citep[BAO; e.g.,][]{eisenstein_2005,anderson_2014}, 
 and most recently, cosmic microwave background (CMB) modeling with \textit{Planck} \citep[e.g.,][]{planck,planck_2015},
  a standard cosmological model has emerged. 
This `Concordance Model' \citep[e.g., ][]{spergel_2007, komatsu_2011}   
 was challenging to envision at the outset of the \ho Key Project \citep{freedman_2001},
  when there existed 10$\sigma$ discrepancies in measurements of \ho 
  for relatively small samples of nearby galaxies \citep{saha_2001}.
At that time, the different programs differed by up to a factor of two in their estimates, 
  although each quoted errors of only 10 per cent on their own measurement \citep[see e.g. Table 6 in][for a summary of the literature at that time]{sandage_2006}.  

\begin{figure*} 
 \centering
\includegraphics[width=1.0\textwidth]{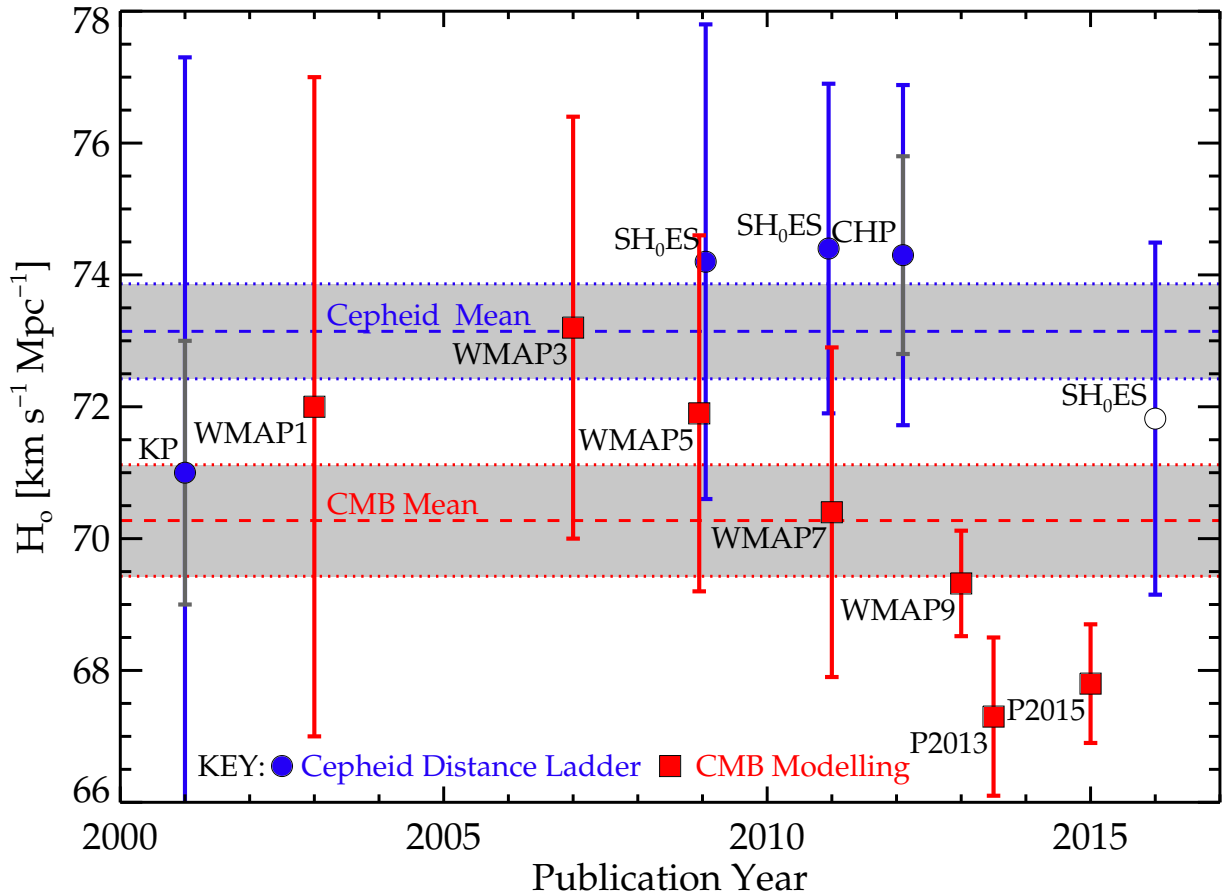}
 \caption{Comparison of recent determinations of \ho using the 
   Cepheid distance ladder (blue circles) and CMB modeling techniques (red squares) as 
   a function of publication year, with the recent value from \citet{riess_2016} given as an open circle.
Many of the projects shown here have multiple measurements, but effort was taken to choose
   the measurements that are most comparable;
   for the Cepheid distance ladder the `preferred value' based on the different anchors is used and
   for the CMB modeling values are shown using similar input priors on cosmological parameters 
    (see Table \ref{tab:ho} for a more exhaustive list of individual and best-estimate values). 
 Colored error bars represent the `total' error (quadrature sum of random and systematic) quoted
   at the time of publication with the grey error bars indicating just the random component
   (but only if it was indicated in the publication; see Table \ref{tab:ho}).
  The 1-$\sigma$ range of the mean for the Cepheid distance ladder and CMB modeling
   are shown as the shaded regions with the mean indicated by a dashed line.
  These two approaches, which use techniques that are `local' and `cosmological' in nature, respectively,
   differ by more than 3-$\sigma$ in their unweighted mean values. 
  As demonstrated here, the tension has grown with the accumulating CMB datasets.
   \label{fig:hotime}}
\end{figure*} 
\begin{table*} 
\begin{center}
\caption{Recent measurements of \ho from CMB Modeling and the Distance Ladder \label{tab:ho}}
\begin{tabular}{l c c c c c} 
 \hline \hline
Short Title$^{1}$ & Technique & Notes & \ho      & Uncertainty & Reference  \\ 
                  &           &       & \hounit  & \hounit     &            \\
 \hline \hline
KP$^{2}$    & Cepheid       & LMC Anchor, \sn          & 71          & 2 (ran), 6 (sys) & \citet{freedman_2001} \\
KP$^{2}$    & Cepheid       & LMC Anchor, Tully-Fisher & 71          & 3 (ran), 7 (sys) & \citet{freedman_2001} \\
KP          & Cepheid       & Preferred Value          & 72          & 8                & \citet{freedman_2001} \\
WMAP1       & CMB           & \wmap One-Year           & 72          & 5                & \citet{spergel_2003}  \\
WMAP3       & CMB           & \wmap Three-Year Mean    & 73.2        & +2.1, -3.2       & \citet{spergel_2007}  \\
WMAP5       & CMB           & \wmap Five-Year Mean     & 71.9        & +2.6, 2.7        & \citet{dunkley_2009}  \\
S$H_{0}$E$^{2}$S  & Cepheid & LMC, Anchor              & 74.2        & 3.6              & \citet{riess_2009}    \\
S$H_{0}$ES$^{2}$  & Cepheid & LMC+MW$\pi$ Anchor       & 74.4        & 2.5              & \citet{riess_2011}    \\
S$H_{0}$ES$^{2}$ & Cepheid  & NGC\,4258, Anchor        & 74.8        & 3.1              & \citet{riess_2011}    \\
S$H_{0}$ES  & Cepheid       & Preferred Value          & 73.8        & 2.4              & \citet{riess_2011}    \\
WMAP7       & CMB           & \wmap Seven-Year Mean    & 70.4        & 2.5              & \citet{komatsu_2011}  \\
WMAP7$^{2}$ & CMB           & \wmap+BAO+\ho Mean       & 70.2        & 1.4              & \citet{komatsu_2011}  \\
CHP         & Cepheid       & LMC Anchor, MIR          & 74.3        & 1.5 (ran), 2.1 (sys) & \citet{freedman_2012} \\
WMAP9       & CMB           & \wmap Nine-Year          & 70.0        & 2.2              & \citet{bennett_2013}  \\
WMAP9$^{2}$ & CMB           & \wmap+eCMB+BAO+\ho       & 69.32       & 0.8              & \citet{bennett_2013}  \\
P2013       & CMB           & Planck+WP+highL          & 67.3        & 1.2              & \citet{planck_2013}   \\
P2015       & CMB           & Planck TT+lowP+lensing   & 67.8        & 0.9              & \citet{planck_2015}   \\
S$H_{0}$ES$^{2}$  & Cepheid & LMC, Anchor              & 71.82       & 2.67             & \citet{riess_2016}    \\
S$H_{0}$ES$^{2}$  & Cepheid & NGC\,4258, Anchor        & 72.02       & 2.51             & \citet{riess_2016}    \\
S$H_{0}$ES  & Cepheid       & Preferred Value          & 73.00       & 1.75             & \citet{riess_2016}    \\
 \hline 
CMB Mean$^{3}$ 		     & CMB       & Unweighted               & 70.3        & 0.8  &  \\
Cepheid Mean$^{3}$       & Cepheid   & Unweighted               & 73.4        & 0.4  &   \\
 \hline \hline
\multicolumn{6}{l}{$^{1}$ Label in Figure \ref{fig:hotime}} \\
\multicolumn{6}{l}{$^{2}$ Omitted from Figure \ref{fig:hotime}} \\
\multicolumn{6}{l}{$^{3}$ Note the mean is computed only for those results displayed in Figure \ref{fig:hotime}.} \\
\end{tabular}
\end{center}
\end{table*} 

Today the situation is qualitatively quite different.  
The local determinations of \ho, using a progressively constructed `distance ladder', 
 traditionally anchored on Cepheid variables, consistently give
 a value of the Hubble constant around 74$\pm$3~\hounit~\citep{riess_2011,freedman_2012,riess_2016}. 
On the other hand, \ho can be estimated from a fit  
 to the measured power spectrum of CMB anisotropies to 
  infer the local expansion rate of the Universe \citep{planck_2013,planck_2015}.
 The CMB value of \ho, extracted from the modeled data, 
  is 67.3 $\pm$ 1.2 \hounit.
 Figure \ref{fig:hotime} presents recent measurements
  of \ho limited to those derived from either the distance ladder (`local')\footnote{For visual simplicity, we have limited this to just those papers providing an end-to-end distance ladder measurement.}
  or the CMB anisotropies (`cosmological').\footnote{Figure \ref{fig:hotime} uses the `preferred' \ho~ results from specific works, which often combine different anchor points for the Leavitt Law. 
We note, however, that while the difference in the distance to the LMC adopted between the KP and the most recent $SH_{0}ES$ determination
 is $<$0.1 per cent, thereby comparisons between those measurements, despite the dramatic improvement of precision of the distance to the LMC \citep[i.e., those adopted in either][]{freedman_2001,pietrzynski_2013}, are meaningful;
 the distance to the megamaser host NGC\,4258, in contrast, changed by $\sim$5 percent between
 \citet{herrnstein_1999} and \citet{humphreys_2013} (used by Riess et al. 2011 and 2016, respectively)
 and, thereby, \ho measurements and uncertainties anchored by either are not immediately comparable 
 without consideration for these changes. 
Additional measurements with different anchors are given in Table \ref{tab:ho} for completeness. 
By demonstrating the impasse between these techniques in Figure \ref{fig:hotime}, 
 we motivate the need for a third method; 
 a detailed comparison between the `best' results from each method is provided in \citet{riess_2016}. }

Figure \ref{fig:hotime} demonstrates that the 
 `local' and `cosmological' solutions are more than 3$\sigma$ apart.
While the distance ladder \ho measurements are consistent even for multiple groups,
 the value of \ho inferred from cosmological modeling, in contrast,
 has monotonically decreased with the acquisition of progressively better CMB maps 
 (in resolution, sensitivity, and spectral coverage).
Once again, an accurate determination of the expansion rate of the Universe comes down
 to an inter-comparison of the stated error bars on the results 
 of two competing methods of measuring \ho~---
 a situation similar to that motivating the HST \ho Key Project twenty years ago \citep{freedman_2001}. 
  
The broad agreement between \ho derived from the distance ladder and 
 from cosmological models is actually quite impressive given 
 the very different methodologies being employed.
However, the value of the Hubble constant applies so much
 leverage to the covariantly-dependent solutions of other cosmological
 parameters (simultaneously embedded in the CMB data) that 
 it remains crucial to determine if there is new physics or simply as-yet-unappreciated systematics
 driving these differences.

\subsection{Uncertainties in Current $H_{0}$ Measurements} \label{ssec:houncertainties}

We turn now to a discussion of the uncertainties in both the distance ladder and CMB methods for determining \ho.
A longstanding `first rung' for the distance ladder is the Large Magellanic Cloud (LMC).
The distance to the LMC is, in itself, 
 a source of active research \citep[see for instance the summary of][]{lmc_distance}
 and its suitability as the ``anchor'' for the distance ladder is much debated \citep[see for instance][]{horevisited}.
The LMC is a low-metallicity, dwarf irregular galaxy,
 unlike the higher-metallicity, higher luminosity spiral galaxies
 that are common hosts to Type Ia supernova (\sn). 
The potential for metallicity effects between their Cepheid populations is therefore a concern.
 Yet, the proximity of the LMC to the Sun permits the application of numerous independent
  distance measurement techniques,
  including those with uncertainties much easier to control for than the Cepheids.
  Of note is the work of \citet{pietrzynski_2013}, 
  in which a set of eight late-type double eclipsing binaries in LMC were used to obtain
  a (largely geometric) 2.2 per cent estimate of the distance to the LMC. 
 On the other hand, the proximity of the LMC also means that its three-dimensional structure
  and line-of-sight depth are both of larger concern than in most extragalactic applications
  and extra attention must be given to membership in specific structural components 
  for a given distance tracer 
  \citep[for further discussion, see][]{clementini_2003, clementini_2011, moretti_2014}.
 Replacement of the LMC with the megamaser-host NGC\,4258 as the anchor point 
 for the distance ladder provides an independent zero-point calibration,
 but does not fully settle the issue of the zero point, given 
  that, effectively, only one dataset exists \citep[with multiple, distinct distance measurements including][among others]{herrnstein_1999, humphreys_2013,riess_2016}, 
  for which the systematic errors are difficult to ascertain.

The advantages of the Leavitt law 
 (or the Cepheid period-luminosity relation\footnote{At the conclusion of the `Thanks to Henrietta Leavitt Symposium' on Nov.~8, 2008, the attendees decided to 
 adopt this nomenclature in honor of her significant contribution to this field. See \url{https://www.cfa.harvard.edu/events/2008/leavitt/} for details.}) 
 are well-known.
(a) Cepheids are intrinsically high-luminosity supergiants.
(b) In the infrared, especially, their period-luminosity relations have 
     small intrinsic dispersion ($\sim$0.1 mag) for the I-band and redder wavelengths.
(c) They are found in all star-forming galaxies (spiral and irregular).
Lastly, (d) their variability is sufficiently stable over a human lifetime
   that they can be repeatedly observed with different instruments operating 
   at different wavelengths and thereby be tested for any number of systematic effects.   
Transient objects (like \sn) are much more problematic in the latter regard. 
But there are remaining challenges for the Cepheid distance ladder, such as: 
 (a) the need for a robust determination of the metallicity dependence of the Leavitt law,
  particularly at optical wavelengths \citep[e.g.,][among others]{romaniello_2005,romaniello_2008},
 (b) the need for additional trigonometric parallaxes to lay a more robust 
  geometric calibration of the zero-point of the Leavitt law \citep[see a prospectus for variables stars in final \gaia sample in][]{eyer_2000},
 (c) contending with crowding at intermediate and large distances,
  especially by redder stellar populations that increase in relative brightness
  moving from the optical into the near- and mid-infrared, 
  and finally,
 (d) potential unknown changes in the reddening law as a function of metallicity and/or environment. 
Some of these concerns were largely minimized by moving into the mid-infrared \citep[e.g., the Carnegie Hubble Program;][see our Figure \ref{tab:ho}]{freedman_2011,freedman_2012}, 
 but tension between the Cepheid distance ladder and other techniques has remained.
 
Only about 800  Cepheids are known in the Galaxy, most of which are located in the Solar neighborhood. 
\gaia is expected to observe up to 9000 classical Cepheids across the Galaxy to a limiting apparent magnitude $G\sim$20 mag \citep{eyer_2000,turon_2012}
 and measure their parallax to better than $\sigma_{\pi}/\pi$ = 10 per cent, on average, 
 and to better than 1 per cent for those located within 1-2 kpc \citep{turon_2012}. 
Beyond the Galaxy, \gaia will measure the parallaxes for at least 1000 of the brightest LMC Cepheids with
  $\sigma_{\pi}/\pi \sim$ 50-100 per cent, providing a probe of the Leavitt law at lower mean metallicity.
In addition to the parallaxes, \gaia obtains metallicity estimates from $R=11500$ spectroscopy 
 with the Radial Velocity Spectrometer (RVS) for all stars brighter than $V\sim16$ mag
 and permits study of star-by-star metallicity effects.
The bulk of the Cepheids observed by \gaia are also accessible for independent, 
 high-resolution abundance studies
 from the ground \citep[e.g.,][among others]{luck_2011}.
In combination, such datasets will allow direct study of the 
 various dependencies of the Leavitt law and permit improvement in the calibration of the Cepheid distance ladder in the future.
 
There are also challenges in estimating \ho from measurements of anisotropies in the CMB. 
The microwave background data from which these measurements are derived are highly complex
 and require highly precise Galactic foreground removal as well as 
 multi-wavelength dust modeling: 
 there is still debate regard the effects on the resulting parameter fits \citep[for instance, see discussion in][]{spergel_2015}.
The power spectrum of CMB fluctuations derived from such maps is then modeled using six primary parameters
 that, hereto, have provided an excellent fit to existing datasets.
However, in addition to the priors imposed on the parameters in the specific fitting procedure employed,
 there exist strong degeneracies in the {\it derived} cosmological parameters 
 and these co-variant degeneracies are strongly coupled to \ho
 \citep[see discussion of such effects in][]{planck_2013}.
Moreover, comprehensive re-analysis by independent teams 
 and detailed comparisons with complementary datasets 
 (often at different frequencies or spatial resolutions)
 can produce significantly differing results \citep[for instance see][]{addison_2015},
 suggesting that there is still more to learn about methodology (both data and analysis) 
 before the results can be interpreted at high confidence.
Moreover, some authors have argued that, with the numerous complementary datasets available, 
 the inclusion of additional parameters in the standard modeling is now feasible and,
  perhaps, more appropriate \citep[][]{divalentino_2015,divalentino_2016}.

\begin{table*} 
\begin{center}
\caption{Summary of Major Observational Programs Referenced in this Work} \label{tab:carnegieprograms}
\begin{tabular}{l c c c c l} 
 \hline \hline
Program Name & Acronym & Primary   & Proposal & Proposal & Distance Indicator\\
             &         & Telescope &          & ID        &      \\
 \hline \hline
  & & & & & \\
\emph{Carnegie Hubble Program}   & CHP  & \emph{Spitzer} & \citet{spitzer_2008} & 60010 & MIR Cephieds \\
 & & & & & \citet{freedman_2012} \\
\hline
\emph{Carnegie RR Lyrae Program} & CRRP & \emph{Spitzer} & \citet{spitzer_2012} & 90002 & MIR RR Lyrae\\
 & & & & & Scowcroft et al. in prep \\
\hline
\emph{Carnegie-Chicago Hubble Program} & \cchp & \hst & \citet{hst_2014} & 13691 (GO)   & NIR RR Lyrae + Optical TRGB \\
                                       &       & \hst & \citet{hst_2013} & 13472 (SNAP) & NIR RR Lyrae \\
& & & & & \\                                      
\hline \hline
\end{tabular}
\end{center}
\end{table*} 

\section{The Carnegie-Chicago Hubble Program} \label{sec:thecchp}

To provide an independent route to \ho and to increase the accuracy of the direct 
  measurements of \ho, the \emph{Carnegie-Chicago Hubble Program} (CCHP), 
  aims to establish a new calibration of \sn through the luminosity of the 
  Tip of the Red Giant Branch \citep[TRGB;][]{lee_1993}, as calibrated, 
  in the first instance, by RR Lyrae stars.
For clarity in placing this new program in the context of other efforts, 
 Table \ref{tab:carnegieprograms} provides a summary of the major Carnegie-led efforts to measure the Hubble constant. 
The TRGB method is extremely precise 
 and the underlying physics is well understood \citep[e.g.,][among others]{salaris_2002}.
Since galaxies of all morphological types host an ancient population of stars,
 the TRGB is a universally-available Population II distance indicator.  
Moreover, RGB stars are found within all galactic structural components, 
 and are the dominant population in their low-density stellar halos. 
Stellar halos convey particular advantages for stellar population studies: 
 (i) they are relatively immune to the vagaries of internal interstellar extinction,
 and (ii) since the stellar density drops with radius \citep[see for instance ][]{gilbert_2012},  
 concerns relating to crowding can be minimized.

 The \cchp~ is designed to take advantage of the TRGB and build a fully independent, 
  high-accuracy, and high-precision determination of \ho 
  based exclusively on Population~II stars. This distance ladder is built using a combination of ground-based observations,
  archival HST observations, and new data acquired 
  in a Cycle 22 GO proposal \citep[Proposal 13691;][]{hst_2014}.
 The program is constructed such that the observations necessary for each rung
  are completed in a uniform fashion (i.e., using the same telescope and instrument combinations)
  to produce datasets designed specifically for the necessary measurements.
 Moreover, the \cchp~ can use archival datasets for its sample of tracer objects
  to test for and calibrate out any systematic effects. 
It is the objective of this paper to give a full overview
  of the steps required to build this independent distance ladder to provide
  context for the individual works to come.

Our distance ladder, built from Population II distance indicators, is currently constructed as follows: 
\begin{enumerate}

\item We first use \hst$+$\,Fine Guidance Sensor (\fgs) trigonometric parallax distances to five Milky Way RR~Lyrae variables (RRL) to set the zero point of the HST near-infrared ($F160W$) RR~Lyrae Period-luminosity (PL) relation on a geometric basis. 
We have obtained new $F160W$ observations for these five stars that will be complemented with 
 data from {\it Spitzer}, data from a ground campaign, and forthcoming measurements from \gaia.
In addition, we are in the process of obtaining improved metallicities of Galactic star clusters to investigate the role of metallicity in the RR~Lyrae PL (Section \ref{sec:rrlyrae}).   

\item We then calibrate the TRGB absolute magnitude by applying the HST near-infrared PL relation to known RRL in the halos of six Local Group galaxies with a seventh calibration object from the megamaser distance to NGC\,4258 (Section \ref{sec:trgbabs}).  

\item Next, we calibrate the absolute luminosity for Type~Ia supernovae (\sn) by applying the TRGB calibration to  determine the distance to their host galaxies (Section \ref{sec:trgbsn1a}).  

\item In the final step, we apply the \sn calibration to the large population of \sn that extend well into the pure Hubble flow by which we measure \ho (Section \ref{sec:hubbleflow}). 
\end{enumerate}

Using reasonable estimates of the errors associated with each step and our sample size,
  this Population II distance ladder can reach an initial precision of 2.9 per cent. 
We discuss individually our random and systematic uncertainties,
 and how they can be improved in the future in Section \ref{sec:discussion}, 
  in particular with the large set of trigonometric parallaxes obtained with \gaia.
A summary is given in Section \ref{sec:summary}.

\begin{figure} 
 \centering
 \includegraphics[width=\columnwidth]{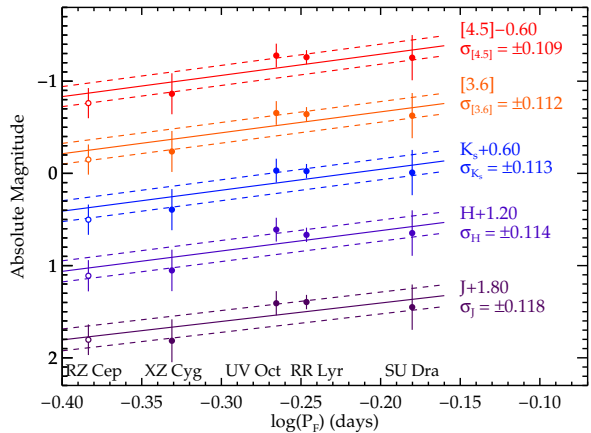}
 \caption{PL relations  
   in the near-- and mid--infrared for the five RR Lyrae variables
   with trigonometric parallaxes with \fgs from \citet{benedict_2011}. 
  There are four RRab type variables (filled circles) and one type RRc (open circle), 
   which is plotted using its fundamentalized period ($P_F$, where $\log(P_{F}) =  \log(P_{FO}) +0.127$). 
  The MIR data comes from the CRRP and the NIR data from 2MASS. 
  Notice the measurable slopes across the wavelength range and low intrinsic scatter ($\sigma_{\lambda} \sim$0.1~mag),
   which is indicated with the dashed lines about each fit.
 \label{fig:zpcal}}
\end{figure} 
\section{The Absolute Scale for Galactic RR Lyrae Variables} \label{sec:rrlyrae}

The first rung of the Population II distance ladder
  is constructed using RR~Lyrae variables in the Galaxy.
RRL are evolved core-helium burning stars that have two primary sub-types,
 the RRab types that pulsate in the fundamental mode (i.e., with pulsational nodes at the core and edge of the star)
 and the RRc types that pulsate in the first overtone mode (i.e., with an additional pulsational node inside the star).
They are typically old with sub-solar masses, and have periods that range 
 from a few hours to one day {\bf (with RRc periods being shorter than RRab periods)}.
Since the horizontal branch is, effectively, flat for some optical wavelengths (at which they were originally observed), 
 the optical period luminosity relationship is less immediately powerful than for Cepheid type variables.
 Distance determinations using the RRL are generally performed  
 using the absolute mean visual magnitude ($M_V$) of the horizontal branch (as determined by the RRL) 
  and the mean metallicity ([Fe/H]) of the population \citep[the $M_V$-Fe/H relation, see][]{cacciari_2003} 
 rather than the direct utilization of the PL itself \citep[for a review of the use of RRL for distance determinations see][]{bono_2003}.
 
In their study of Galactic globular clusters at 2.2$\mu$m, \citet{longmore_1986}
 first noted a correlation between period 
 and the corresponding single phase 2.2 $\mu$m magnitude for the RRL
 and suggested that there might exist a near--infrared (NIR) PL.
Later, \citet{longmore_1990} used this first NIR PL 
 to measure the distance to eight Galactic globular clusters in the $K$ band.
Extensive theoretical modeling demonstrated  
 how the wavelength-dependent bolometric correction ultimately drives the behavior
 of the PL from the optical and into the infrared \citep{bono_2001,bono_2003,catelan_2004,marconi_2015}.
Moreover, the very tight mass--luminosity relationship for the horizontal branch
 phase of stellar evolution forces the PL to have a very small intrinsic 
 width at these wavelengths \citep[see][]{catelan_2004}.
Thus, the use of the NIR PL in application to distances is particularly advantageous.

Several other significant advantages exist for observations of variable stars undertaken in the NIR:
\begin{enumerate}
\item the reduction of the effects of line-of-sight extinction by almost an order of magnitude \citep[$A_H$/$A_V$ = 0.178, see][]{cardelli_1989}; 
\item the lower systematic impact of the possible non-universality of the reddening law; 
\item the total amplitude of the light variation of the target star, seen in the near-infrared during its pulsation cycle, is greatly reduced due to the diminished contribution of temperature variation to NIR surface brightness \citep[e.g.,][among others]{longmore_1990,jones_1996};
\item the sinusoidal shape of the light curve due to the dominance by the radial variation effects in the NIR \citep[see for instance][among others]{jones_1988,skillen_1993,smith_1995};
\item the corresponding reduction in the intrinsic scatter of the PL relations \citep{madore_2012}, again because of the reduced sensitivity of infrared luminosities to temperature variations; and finally, 
\item at the temperatures and surface gravities encountered for RR Lyrae variables, there are few metallic line or molecular transitions in the NIR that atmospheric metallicity effects are predicted to be low. The CO band-head within the IRAC 4.5 $\mu$m filter known to affect Cepheid light curves \citep[e.g.,][]{monson_2012, scowcroft_2016} is not anticipated to have a significant effect for these temperatures. Furthermore, any remaining effects can be tested for and calibrated out.
\end{enumerate}

Several recent and large-scale programs have begun to exploit the NIR and MIR PL,
 including the \emph{Carnegie RR~Lyrae Program} \citep[CRRP;][]{spitzer_2012}, 
 the \emph{Spitzer Merger History and Assembly of the Stellar Halo} \citep[SMHASH;][]{spitzer_2013},
 the \emph{VISTA Variables in the Via Lactea} project \citep[VVV;][]{vistavariables}, and 
 the \emph{VISTA near-infrared $YJK_s$ Survey of the Magellanic Cloud System} \citep[VMC;][]{cioni_2011}.
In this program, we will use the NIR PL as the foundation for a Population II distance ladder.
Our first step is the precise calibration of the NIR PL for the IR channel of \hst$+WFC3$;
 in particular, we use the $F160W$ filter, which is similar to the more familiar ground-based $H$. 

There are two primary concerns that are addressed at this stage:
  (i) the absolute $F160W$ luminosity, i.e., the zero-point of the PL relation (Section \ref{ssec:trigparallax}), and 
  (ii) and evaluation of scatter due to astrophysical effects 
    \textit{within} a population of RR~Lyrae (Section \ref{ssec:rrlmetals}).
 We utilize a set of Galactic RR~Lyrae with pre-existing parallaxes for the former;
 whereas for the latter, we explore a subset of RR~Lyrae variables
  in the \ocen globular cluster. 
\gaia will deliver parallaxes with about 10 microarcsecond uncertainty for the 100-150 brightest
 RR~Lyrae \citep[those with $V$\textless $\sim$12.5 mag;][]{debruijne_2014,clementini_2016}
 and will transform our understanding of both the absolute zero-point and 
 most effective method to account for astrophysical effects.
For the purpose of this paper, we focus discussion to the current attainable precision 
 for NIR observations of RR Lyrae and discuss sources of astrophysical uncertainty
 as they relate to our projected error budget, 
 which is given in Section \ref{ssec:rrlerror}. 

\begin{table*} 
\begin{center}
\caption{Properties of RR~Lyrae variables in the \cchp~ Galactic calibration sample. \label{tab:rrl_calibrators} }
\begin{tabular}{l c c c c c c c c c } 
 \hline \hline
Target & Type & $\log(P_{F})$ & [Fe/H]$^{1}$ & $\langle V\rangle^{1}$ & ${\bf \langle H\rangle^{2}}$ & $A_{V}^{3}$ & $A_{H}^{4}$ & $\pi^{1}$ & $\mu$           \\ 
 \hline \hline
SU Dra & RRab & -0.180       & $-1.80 \pm 0.20$ & 9.78 & 8.69 $\pm$ 0.03 & 0.03 & 0.01 & $1.42 \pm 0.16$ & $9.28 \pm 0.24$ \\
RR Lyr & RRab & -0.247       & $-1.41 \pm 0.13$ & 7.76 & 6.61 $\pm$ 0.01 & 0.13 & 0.02 & $3.77 \pm 0.13$ & $7.11 \pm 0.07$ \\
UV Oct & RRab & -0.266       & $-1.74 \pm 0.11$ & 9.50 & 8.30 $\pm$ 0.02 & 0.28 & 0.05 & $1.71 \pm 0.10$ & $8.84 \pm 0.13$ \\
XZ Cyg & RRab & -0.331       & $-1.44 \pm 0.20$ & 9.68 & 8.79 $\pm$ 0.04 & 0.30 & 0.05 & $1.67 \pm 0.17$ & $8.89 \pm 0.22$ \\ 
RZ Cep & RRc  & -0.384$^{5}$ & $-1.77 \pm 0.20$ & 9.47 & 8.06 $\pm$ 0.05 & 0.75 & 0.13 & $2.54 \pm 0.19$ & $7.98 \pm 0.16$ \\
 \hline \hline
\multicolumn{9}{l}{$^{1}$ From \citet{benedict_2011}          } \\
\multicolumn{9}{l}{{\bf $^{2}$ Derived single \emph{2MASS} phase points following the technique described in Section \ref{ssec:templates}.}}\\
\multicolumn{9}{l}{$^{3}$ From \citet{feast_2008}             } \\
\multicolumn{9}{l}{$^{4}$ Assuming $A_{H}/A_{V}=0.178$        } \\
\multicolumn{9}{l}{$^{5}$ $\log(P_{F}) =  \log(P_{FO}) +0.127$}
\end{tabular}
\end{center}
\end{table*} 

\subsection{The Current Geometric Foundation of the RR Lyrae Distance Ladder} \label{ssec:trigparallax}

 Trigonometric parallaxes have been obtained for five Galactic
  RRL using the \hst Fine Guidance Sensor \citep[\fgs;][]{benedict_2011}.
 The \hip~ measurements for RR Lyrae were at its measurement limit;
  with the exception of RR Lyr ($\sigma_{\pi}/\pi \sim$ 18 per cent), the over one hundred
  RR Lyrae in \hip~ have parallax uncertainties 
  greater than 30 per cent \citep[with over 100 stars in the sample; see][]{clementini_2016}.
 Practical constraints limited the application of \fgs to only the most nearby RR Lyrae. 
 As such, trigonometric parallaxes were obtained for only five stars, 
  of which four were type RRab  (SU Dra, RR Lyr, UV Oct, and XZ Cyg)
  and one type RRc (RZ Cep).
 Properties of these five stars are given in Table \ref{tab:rrl_calibrators}. 
 
\subsubsection{Uncertainty in the Period-Luminosity Zero Point} \label{ssec:plzp}

 In Figure \ref{fig:zpcal}, we combine the \citet{benedict_2011} trigonometric parallaxes
  with NIR photometry from 2MASS \citep{skrutskie_2006} and mid--infrared (MIR) photometry from the CRRP
  to demonstrate the NIR ($J$, $H$, $K_s$) and MIR (\emph{Spitzer} 3.6~$\mu$m and 4.5~$\mu$m) PLs.
 The five stars give sparse sampling in $\log(P_{F})$\footnote{Where $P_{F}$ is the fundamental period, 
  for which the first-overtone pulsators 
  are fundamentalized by the expression $\log(P_{F}) =  \log(P_{FO}) +0.127$.}
  and it is not feasible to determine the slope directly from the parallax sample. 
 Thus, we adopt the PL slopes measured in M\,4 by \citet[][their table 2]{braga_2015} 
  for the $J$ (-1.739 $\pm$ 0.109), $H$ (-2.408 $\pm$ 0.082), and $K_s$ (-2.326 $\pm$ 0.074)
  and by \citet[][their table 3]{neeley_2015} for 3.6~$\mu$m (-2.332 $\pm$ 0.106) and 4.5~$\mu$m (-2.336 $\pm$ 0.105).
  
 The adopted PL is over-plotted in Figure \ref{fig:zpcal} for each passband.
 We estimate the impact on our measurements from the uncertainty on the slope by performing
   a Monte-Carlo simulation varying the slope within its uncertainties and find a modulation 
   of the resulting zero-point on order of 0.002 mag or smaller for all bands. 
 Because we define the zero-point of the PL relation
   at the mid-point of our $\log(P)$ distribution (here, $\log(P_F)$ = -0.30),
   the effect of the slope uncertainty on the zero point is effectively eliminated. 
 Moreover, we test the full range of metallicity slopes in \citet[see][their table 3 for a summary]{muraveva_2015} 
  and find no significant impact on the zero-point 
 (defined at the mid-point of [Fe/H] = -1.58 dex)
  for metallicity terms between 0.235 mag dex$^{-1}$ and 0.03 mag dex$^{-1}$. 
 We suspect the lack of an effect of metallicity is 
 due to the lack of a large metallicity span 
 in the \hst calibrators and that they are relatively clustered (see our Table \ref{tab:rrl_calibrators}).
 Thus, we neglect the metallicity term for this demonstration owing to
  both the large range of values for the effect in the literature, 
  that empirical studies are, largely, limited to the $K_{s}$ band at this time,
  and that we see no strong impact for our current RRL calibration sample
   (the individual uncertainties are much larger than the differential metallicity effect).

 As shown in Figure \ref{fig:zpcal}, 
  the variance about the PL for all pass-bands is approximately $\sigma_{\lambda} \sim$ 0.11, 
  though the mean uncertainty on the distance moduli is $\langle \sigma_{\mu} \rangle = $0.15~mag 
  (seven times larger than the mean apparent magnitude uncertainty 
  of $\langle \sigma_{H} \rangle$ = 0.03 mag; Table \ref{tab:rrl_calibrators}). 
 As anticipated by the dominance of the parallax uncertainty to the total uncertainty 
  in the absolute magnitude, the residuals for each band are highly correlated. 
 Indeed, for RRc variables, the only independent measurement of the absolute magnitude
  by \citet{kollmeier_2013} is in 2.5$\sigma$ tension with the \citet{benedict_2011}
  value for RZ Cep. 
 This highlights the critical need for accurate and precise absolute magnitude measurements
  for RRL.
 
 We compute the $\chi^{2}$ statistic compared to the PL for each of our passbands,
  finding $\chi^{2}$ = 0.39, 0.40, 0.42, 0.38, and 0.36 
  for the $J$, $H$, $K_s$, $[3.6]$, and $[4.5]$, respectively.
 For the purposes of the following discussion we use $\chi^{2}_{H} = 0.40$ 
  or the reduced chi-squared metric of 
  $\chi^{2}_{H} \nu^{-1}$ = 0.1 (where $\nu$ is the number of degrees of freedom), 
  which is representative of results for all bands.
 For $\chi^{2} \nu^{-1} < 1$, the implications either are 
  (i) the model is incorrect, 
  (ii) the uncertainties follow a non-Gaussian distribution,
  (iii) improper assignment of uncertainties \citep[i.e.,][chapter 11]{bevington_2003}, or
  (iv) the failure to account for the intrinsic variance of a sample 
   \citep[i.e.,][their section 3.1]{bedregal_2006}.
 We can evaluate each of these possibilities. 
 
 For the first option, our adopted PL is supported by several studies of the PL for globular clusters \citep[][being the most recent]{braga_2015,neeley_2015} and,
  thus, our model is representative of the behavior of NIR and MIR RR Lyrae.
 We did neglect metallicity, but accounting for this would only {\it reduce} the measured
  variance and, using the maximal estimation for the metallicity component \citet[][in the $K$ band]{bono_2001},
  we obtain an effect of 0.07 mag, which, once removed,
  only lowers the measured variance to 0.08 mag.
 For the second option regarding the nature of the uncertainties, 
  discussions involving the uncertainty for the reference frame
  in \citet[][their figure 4]{benedict_2011},
  imply that the final uncertainties are, indeed, Gaussian in nature using their procedure. 
 For the third possibility, we utilize the analysis of \citet[][their figure 7]{benedict_2002}, 
  in which the results for the early \fgs parallax program were compared to \hip,
  from which the authors conclude it is likely that the uncertainties for 
  the \fgs parallaxes are overestimated by a factor of $\sim$1.5.
 Comparing \citet{benedict_2002} to \citet{benedict_2011}, 
  we find no significant change of procedure
  that would have altered the conclusion reached by this comparison.
 
 For the fourth possibility, we undertake a series of Monte-Carlo simulations,
  described in detail in a forthcoming publication focused on the zero-points,
  to understand the interplay between the intrinsic dispersion about the PL
  and the parallax uncertainties in a statistical sense.
 These simulations use a well measured PL as a fiducial by which to 
  compare the parallax uncertainties. 
 Even with the intrinsic variance of the PL included,
  the resulting variance of the fits is still statistically 
  inconsistent with the input uncertainties (i.e., $\chi^{2}_{H} \nu^{-1}$ \textless 1).
 If the parallax errors are scaled by 50 per cent, our simulations produce
  final variances consistent with those measured in Figure \ref{fig:zpcal}
  with $\chi^{2}_{H} \nu^{-1} \sim$ 1.
 Thus, comparison of the \fgs parallaxes to two independent fiducial measurements, 
  \hip and the well established NIR PL,
  indicate that the uncertainties are overestimated by $\sim$50 per cent.
 
 Our assessment is consistent with prior works utilizing the \fgs parallax program 
  for distance scale applications using both Cepheids and RR Lyrae 
  \citep[for instance,][among others]{freedman_2012,neeley_2015,braga_2015, muraveva_2015, clementini_2016}.
 For the purpose of forecasting an error budget for the \cchp, 
  we scale the dispersion, $\sigma = 0.11$ mag, about the PL 
  shown in Figure \ref{fig:zpcal} by 50 per cent, to project that $\sigma = 0.05$ mag
  is statistically consistent with current data.
 While these statistical arguments are compelling, they are not conclusive.
 Owing to the release of initial parallaxes for objects common to \hip
  in the near future \citep[see][]{michalik_2015}, 
  a more exhaustive exploration of the zero-point for the NIR+MIR PL
  (and perhaps a better understanding of the Benedict et al.~uncertainties)
  will feasible at a later stage of our program.
 
\begin{figure*} 
\centering
 \includegraphics[width=\columnwidth]{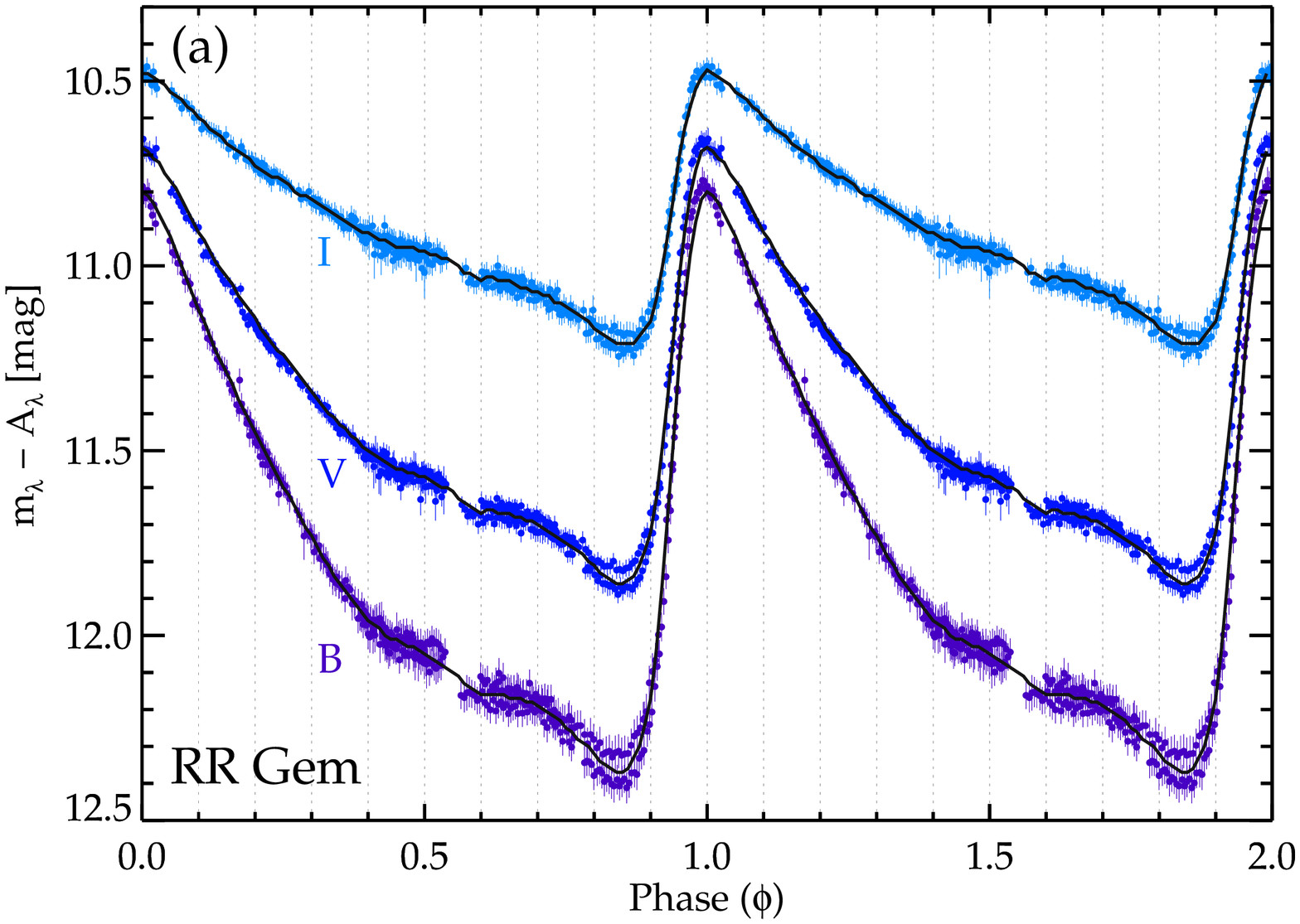}
 \includegraphics[width=\columnwidth]{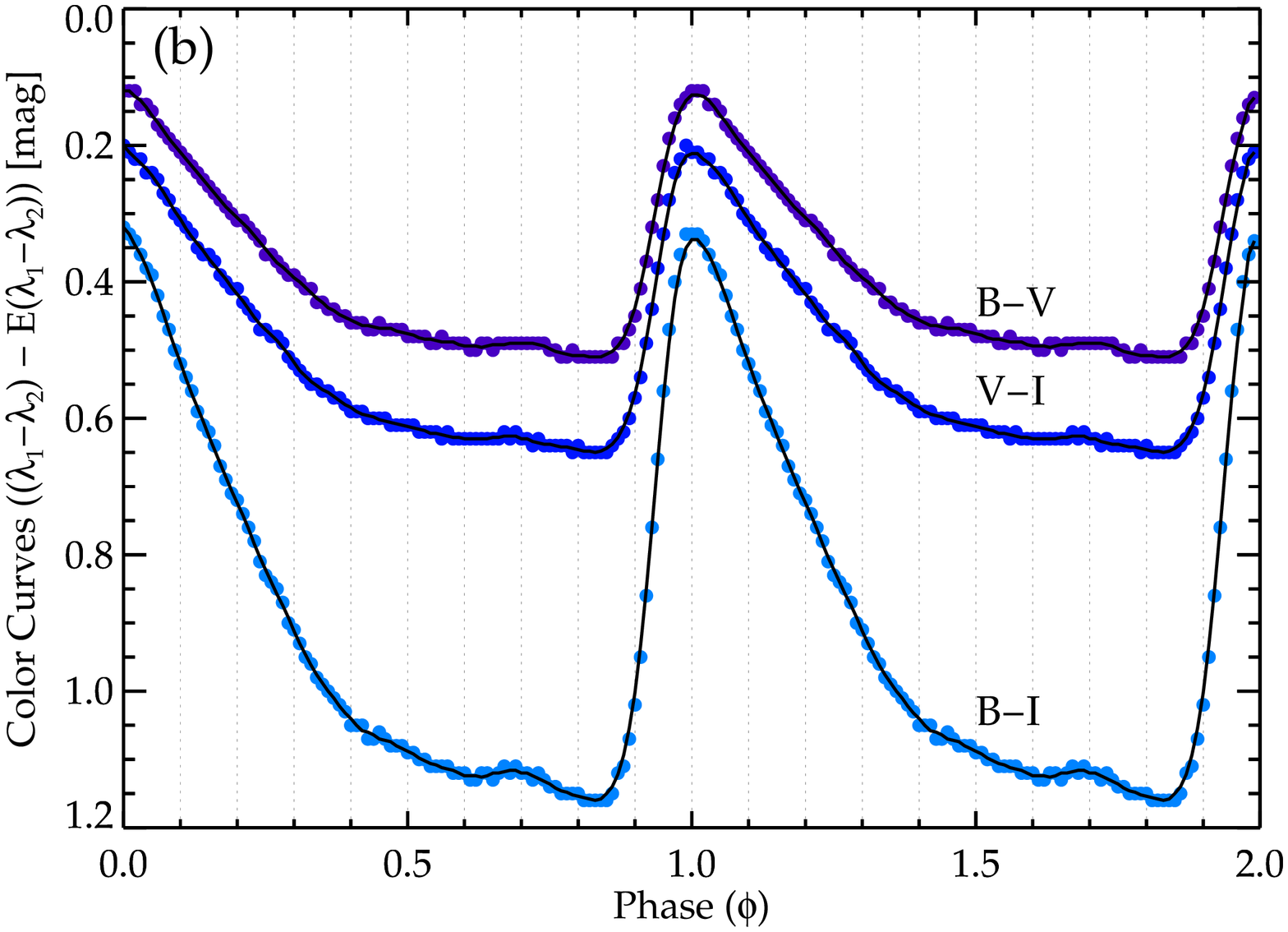}
 \includegraphics[width=\columnwidth]{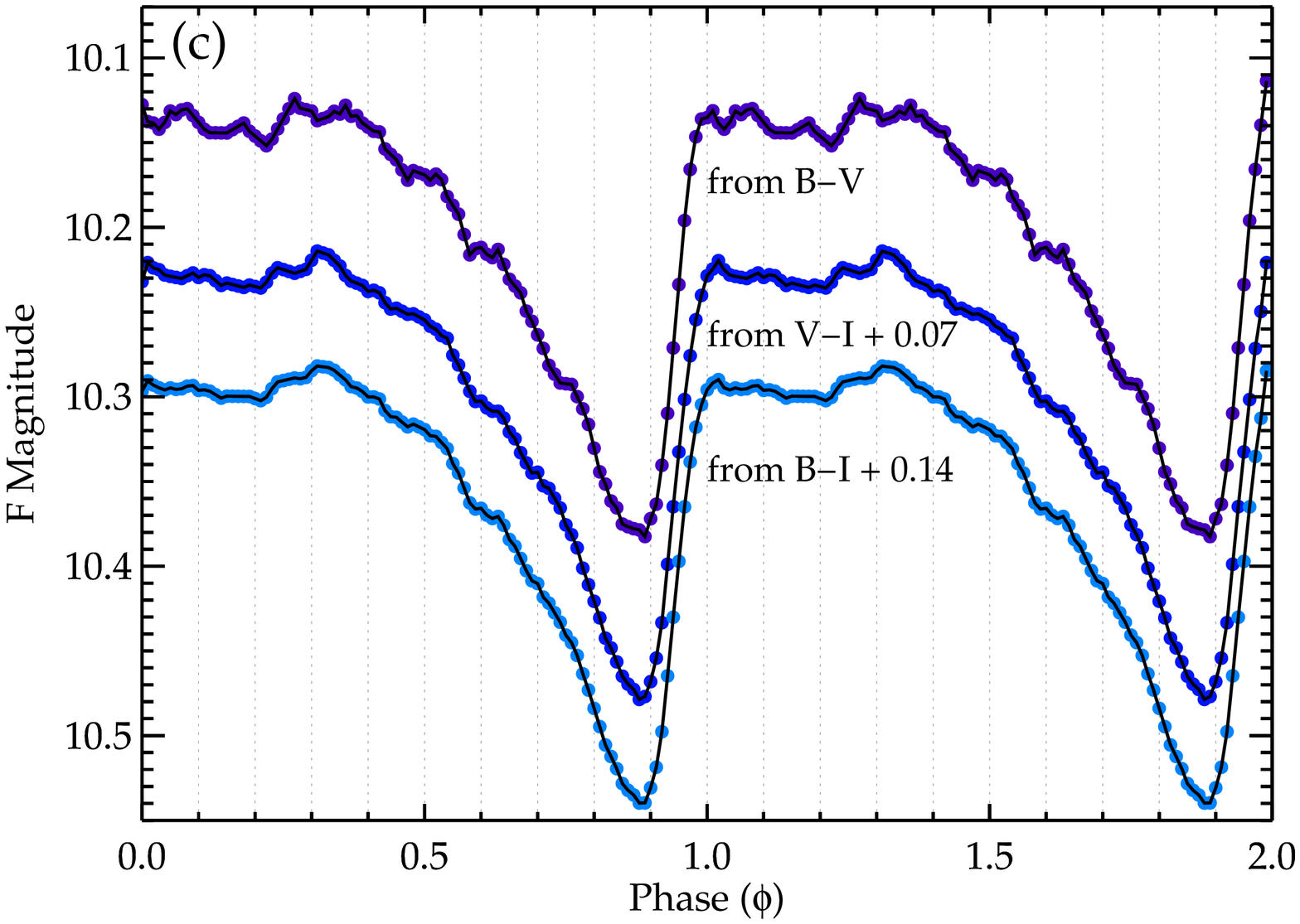}
 \includegraphics[width=\columnwidth]{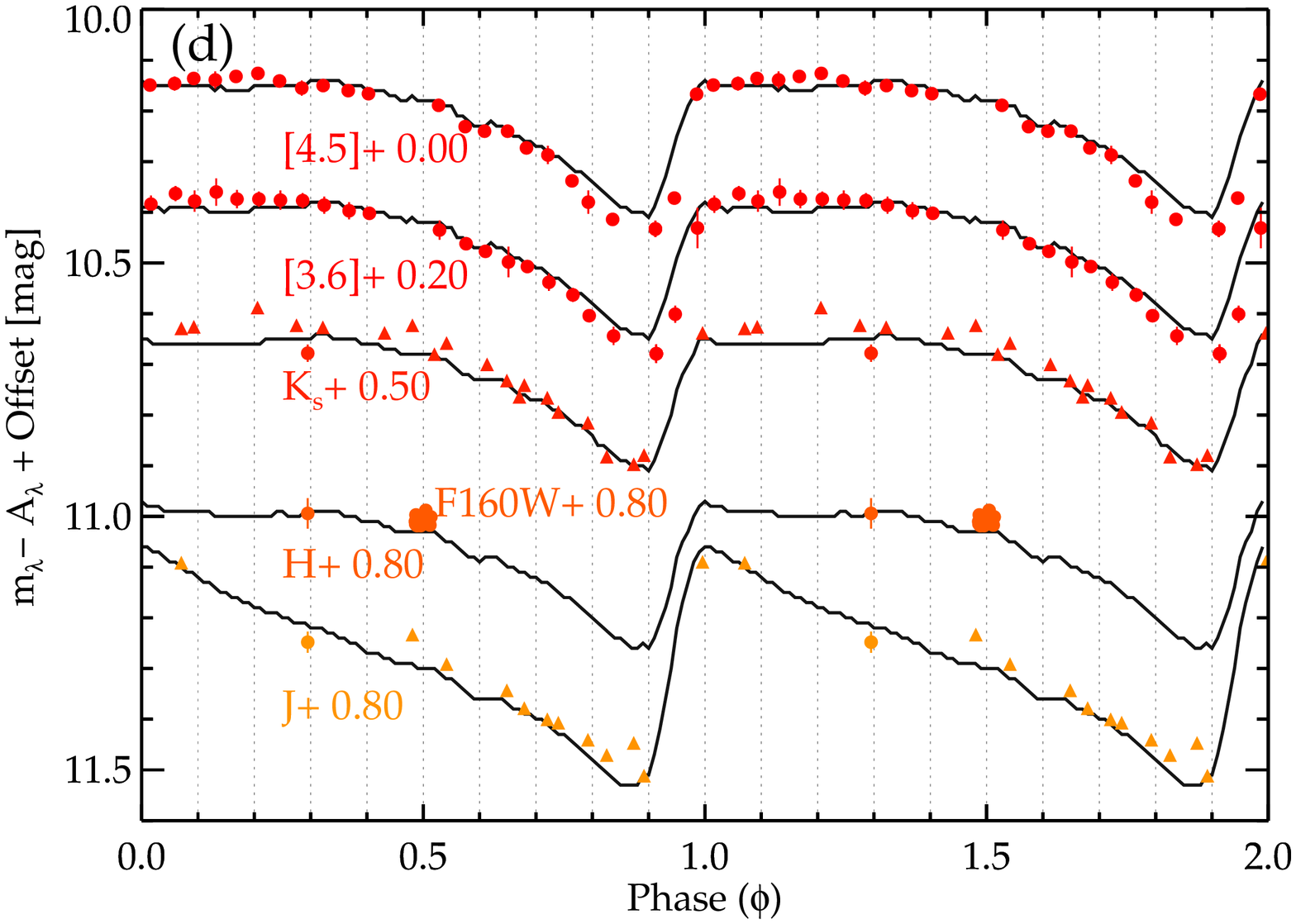}
\caption{Construction of light-curve templates. 
  {\bf (a):} Observed optical light curves for RR Gem in $B$, $V$, $I$.
  {\bf (b):} Color curves for RR Gem that physically trace the temperature variations over the pulsation cycle.
  {\bf (c):} By removing the temperature curves from the initial light curves, the radial variations can be isolated. 
    These radial variations are the only dominant effect seen in the NIR and MIR.
  {\bf (d):} NIR and MIR light curve templates generated from the optical data (black lines) 
    with sparsely sampled MIR data over-plotted (colored points). 
    For $J$ and $K_{s}$ we show both the \emph{2MASS} data (filled circles)
     and data from \citet{liu_1989} (filled triangles).
    For $F160W$ (similar to $H$), we show all of the individual measurements obtained 
     using our \hst observing strategy for this bright star (described in Section \ref{ssec:rrlobserving}).
    Note the reduced light curve amplitudes and overall more sinusoidal shape at these wavelengths.
  Using the templates, we can produce high quality intensity-weighted mean magnitudes even with only sparsely 
   sampled or single phase observations.
  \label{fig:template}}
\end{figure*} 

\subsubsection{HST Observations of Galactic RR Lyrae} \label{ssec:rrlobserving}

We have obtained randomly-phased observations of the five galactic RR-Lyrae with known geometric parallaxes 
 with \hst$+WFC3/IR$ in the $F160W$ filter in our Cycle 22 GO proposal \citep[5 stars; Proposal 13691;][]{hst_2014}. 
Given the relatively bright apparent $H$ magnitudes of these stars,
 ranging from $\langle H \rangle=$ 6.67 to $\langle H \rangle=$ 8.79
 (Table \ref{tab:rrl_calibrators}), 
 special observing techniques were required that are sufficiently technical 
 to be beyond the scope of this work. 
 A full presentation of this stage of our program will be given in Rich et al.~(in prep.).

Obtaining even a single phase point requires a full \hst orbit
 and attaining full phase coverage of these stars is prohibitively expensive
 (i.e., $\sim$12 individual points corresponds to $\sim$12 orbits per star).
Thus, we are limited to single phase points for these stars.
In the next section, we discuss a method to determine accurate mean magnitudes from single phase observations,
 using existing high-cadence light curves for at least two optical wavelengths.
 

\subsubsection{RR~Lyrae Light Curve Fitting for Predictive Templates} \label{ssec:templates}

 To estimate the mean magnitudes of our Galactic RRL calibrators directly from the $F160W$ observations,
   we employ a technique that uses well-sampled optical light curves
  to generate a template suitable for longer wavelength data.
Instead of using Fourier fits modeled to high-cadence NIR observations for select stars
  \citep[i.e., as in][]{jones_1996},
  we will use high-cadence observations in multiple optical bands to derive
  NIR and MIR light curve templates for each individual star {\it from its own optical light curve}.
 This technique has the advantage capturing the specific color-structure of an individual RR Lyrae's light curve.
 
 Our template technique was first developed for Cepheids by \citet{freedman_1988} and \citet{freedman_2010b},
  and requires high cadence optical ($B$, $V$, $I$) time series data. 
 We have initiated an observational survey to obtain such data and derive templates
  for individual stars 
  that will be fully described in a series of forthcoming publications
  (the optical data, not including the new \hst observations, in Monson et al. in prep and the template technique in Beaton et al. in prep).
 To demonstrate the technique and its role in the \cchp, 
  we present data from our optical campaign for RR Gem of RRab type in Figure \ref{fig:template}. 

 The optical light curves for RR Gem are given in Figure \ref{fig:template}a for 
  the $B$, $V$ and $I$ broadband filters.
 These optical luminosity variations
  are sensitive to two physical effects that change due to the pulsations: 
  (i) temperature and (ii) radius.
 The temperature variability can thus be isolated using the optical color 
  as a proxy for temperature.
 The optical color curves for RR Gem are given in Figure \ref{fig:template}b for 
  $(B-V)_{0}$, $(V-I)_{0}$, and $(B-I)_{0}$.
  
 With the temperature effect constrained by color curves, it is then possible
  to remove it from the optical light curves and, thereby, isolate the 
   component due to radial variations. 
 The residual curves after a scaled subtraction of the color curves, 
  called the $F$ magnitude by \citet{freedman_2010b},
  are given in Figure \ref{fig:template}c for each of the color curves 
  in Figure \ref{fig:template}b.   
 These $F$ curves predict the shapes of the NIR and MIR light curves,
  modulo scale factors \citep[see discussion in][]{freedman_2010b}.
 Figure \ref{fig:template}d compares the NIR and MIR light curves predicted 
  from the $F$ magnitude to $K_s$ and $J$ photometry from \citet{liu_1989}
  and MIR photometry from the CRRP,
  which show very good overall agreement.
   
  We can compute the shift between our template light curve in a given photometric band 
   and the single phase point to adjust the photometric zero point of our template,
   as is demonstrated for the 2MASS photometry of RR Gem in Figure \ref{fig:template}d. 
  With the zero point shift, 
  the template can then be used to determine the appropriate mean intensity in a given band.
  Thus, from a single phase point for the \cchp~ in the $F160W$ filter,
   we can determine reliable mean magnitudes.
  Example preliminary measurements from our HST SNAP proposal \citep[PID 13472;][]{hst_2013} 
   in $F160W$ are also plotted in Figure \ref{fig:template}d for RR Gem
   and show overall agreement with the ground-based $H$ curve
 (we note that RR Gem had few of the complications of the significantly brighter 
    \hst RRL calibrators and was chosen as a demonstration object for this reason).
 
The application of this method requires knowledge of RRL ephemerides 
   (time of maximum light, period, and any period-change related terms) to high precision.
  Our optical campaign is sufficiently well sampled
   that when tied into literature observations we are able to test and revise
   the ephemerides for our RRL calibrator stars and be relatively 
   confident in our phasing (Monson et al. in prep).
  The uncertainty in the mean magnitude is a function of phase ($\phi$)
   and will be discussed in Beaton et al. (in prep).
  The mean $H$ magnitudes and uncertainties for the RRL calibrators in Table \ref{tab:rrl_calibrators}
   and used in Figure \ref{fig:zpcal} were determined from a single 2MASS observation
   from this technique.

\begin{figure*} 
 \centering
 \includegraphics[width=\linewidth]{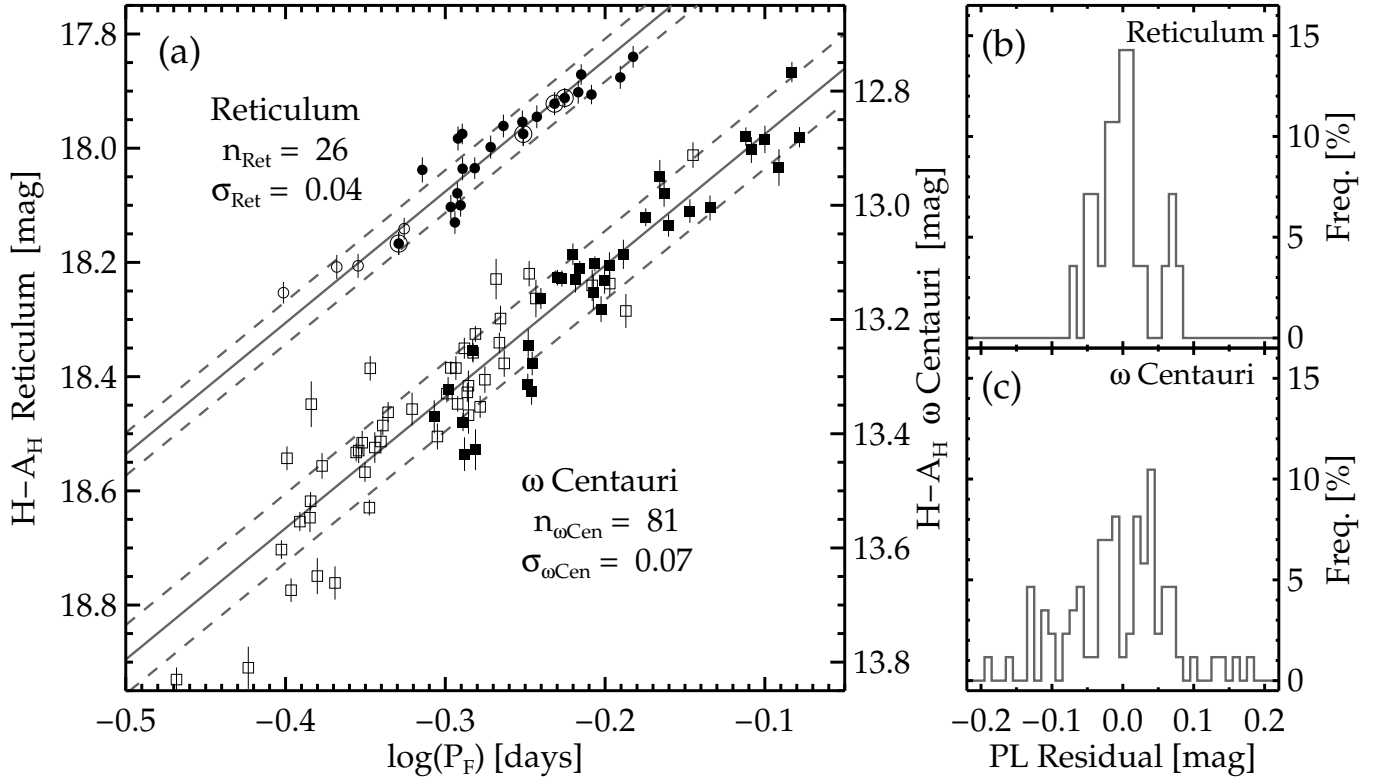}
 \caption{Scatter about the $H$ PL.
 {\bf(a)} Near-infrared ($H$) Period-Luminosity relations for
   26 RR~Lyrae variables in the Reticulum cluster 
   with a scatter of only $\pm$0.04~mag (upper) and
   for the 81 least crowded RR~Lyrae variables in
   \ocen, showing a scatter of only $\pm$0.07~mag (lower).
  In the plot, the open symbols denote RRc type variables, filled denote RRab type variables.
  There are four RRab pulsators with the Blashko effect in Reticulum
   indicated by a larger open circle.
  Solid lines show the fitted PL relationships based on the slope 
   determined from the Galactic globular cluster, M\,4 \citep{braga_2015}, 
   and dashed lines show the $\pm$1-$\sigma$ ranges about the PL.
  This inter-comparison places an upper limit on the contribution of 
   effects that are astrophysical in nature (metallicity, evolutionary state, etc.)
   to the observed scatter in \ocen.
 {\bf(b)} Marginal distribution of PL residuals for Reticulum with a scatter of $\sigma=0.04$ mag.
 {\bf(c)} Marginal distribution of PL residuals for \ocen with a scatter of $\sigma=0.07$ mag.
 \label{fig:metals}}
\end{figure*} 

\subsection{Assessing the Scatter About the PL} \label{ssec:rrlmetals}

Our key concern at this state of the \cchp~ is the precision 
 to which we can measure the distance to our Local Group targets 
 (see Table \ref{tab:rrltargets}, bottom section). 
Our Local Group targets span a range of [Fe/H] from -1.68 dex (Sculptor) to -0.44 dex (M\,32),
 with mean and median metallicity of -1.3 dex and -1.5 dex, respectively 
 (values based on the $M_{V}$-[Fe/H] relationship for the RRL in our pointings are given in Table 4, unless otherwise noted).
 Thus, our Local Group sample is relatively well matched to the mean metallicity of our
  five RRL calibrators (median$_{[Fe/H]}$= -1.58 dex).
Our distance precision is limited by sources of scatter that are either observational (i.e., measurement) 
 or astrophysical in nature. 
For the former, typical effects are photometric accuracy, repeatability, and crowding,
 which generally can be identified and controlled for an individual dataset. 
For the latter, there are numerous physical factors that add scatter to the PL,
 including but not limited to, (i) chemical abundances, (ii) the evolutionary state,
  and (iii) changes in period \citep[e.g.,][]{smith_1995,cacciari_2013}.
 
Diagnosis of such astrophysical complications require independent datasets 
 or specialized statistical methods to be properly taken into account
 \citep[see discussion in][their appendix B]{sandage_2015}.
Unfortunately, for our Local Group targets such independent measurements are prohibitively 
 expensive for all of the six objects;
  the mean magnitude of an RR Lyrae at M31 distance is $V \sim 25.5$ mag,
  which is out of reach for current high resolution spectroscopy
  and requires \hst or imaging on an \textgreater8 meter facility for high quality temporal monitoring.
Thus, to understand the role of these astrophysical effects on the precision of our 
 Population II distance scale, we can only account for them in a statistical fashion.
With comparison of the scatter about a local, 
 well-characterized `ideal' PL to a local, well-characterized analogue to our Local Group RR Lyrae targets,
 it is possible to make such a measurement.
  
To assess the scatter about the PL for our NIR data,
 we compare the $H$ PL for the LMC star cluster Reticulum
  to that of the star cluster \ocen. 
Reticulum is a cluster of 
 (i) low central concentration \citep{walker_1992}, 
 (ii) mono-abundance \citep[$\text{[Fe/H]}=-1.57 \pm 0.03$;][]{grocholski_2006}, 
 (iii) an apparent single stellar population \citep{mackey_2004}, 
 (iv) located $\sim 11^{\circ}$ from the LMC center \citep{walker_1992}, and 
 (v) low apparent fore- and back-ground contamination \citep{mackey_2004}.
In contrast, \ocen is one of the most massive and metal-rich star clusters  
 with a wide metallicity spread for its RR Lyrae \citep[$-2.5 <\text{[Fe/H]}< -1.1$;][]{rey_2000, sollima_2008}
 and a rather complex horizontal branch morphology \citep[for discussion see][]{gratton_1986}.
Moreover, while we have `recent' NIR light curves that can be used to correct for
 shifts in period for individual stars in Reticulum (a detailed description of how this is done 
  in application to our Galactic RRL calibrators will be given in Monson et al., in prep), 
 we do not made such changes in \ocen. 
The median [Fe/H] for Reticulum and \ocen are well matched to the median
 of our Local Group sample.
 Moreover, the total range of metallicity in \ocen is also 
 comparable to the total range of our Local Group sample,
  albeit more metal poor overall (see Table \ref{tab:rrltargets}).
Thus, a comparison of the PL scatter between these two systems provides a meaningful
 estimate of the added uncertainty due to astrophysical scatter in the RRL populations
 we anticipate for our Local Group sample.

Figure \ref{fig:metals}a compares the $H$ PL for Reticulum and \ocen using
 image data collected with the FourStar camera on the Baade 6.5m telescope
 at Las Campanas Observatory \citep{persson_2013} 
 and mean magnitudes determined from {\tt GLOESS} fitting \citep[see discussion in][]{persson_2004}. 
The data for both clusters are overall of the same quality
 (i.e., despite the larger distance to Reticulum, the same signal-to-noise is reached).
As in Section \ref{ssec:plzp}, we adopt an $H$ period slope from \citet[][]{braga_2015}
  for the fundamental period ($P_F$).
The study of \citet{sollima_2006} for 16 star clusters found no significant difference in the slopes for clusters spanning a range of nearly 2 dex in [Fe/H] -- a very similar span as for \ocen (see Table \ref{fig:metals}).
Since the goal of this demonstration is to study the scatter and not the zero-point itself, 
 we neglect a correction for the mean metallicity for either system because
 it only modulates the absolute magnitude of the zero point \citep{muraveva_2015,marconi_2015}. 
We limit the \ocen sample to those stars outside of the \ocen 
 half light radius \citep[][i.e., $r$\textless$5'$]{trager_1995}\footnote{As presented in \citet{harris_1996,harris_1996cat} and updated online versions of the compilation.}
 to avoid complications from crowding in its very dense inner regions.
We are left with a sample of 81 stars in \ocen and 26 stars in Reticulum. 

We determine residuals about the PL as the difference between the magnitude of a given source
 and its magnitude were it on the PL relation (i.e, vertical from the PL in Figure \ref{fig:metals}a). 
A normalized marginal distribution for the residuals is shown in Figure \ref{fig:metals}b 
 for Reticulum and Figure \ref{fig:metals}c for \ocen.
From comparison of Figures \ref{fig:metals}b and Figures \ref{fig:metals}c it is clear
 that the residuals for \ocen~ show a much larger overall variance than those of Reticulum.
For Reticulum, we measure a scatter of $\sigma_{tot} = 0.04$ mag with a median $H$ magnitude 
 uncertainty of $\sigma_{meas} = 0.02$ mag 
 and for \ocen, we measure $\sigma_{tot} = 0.07$ mag with a median $H$ magnitude 
 error of $\sigma_{meas} = 0.02$ mag.
We now use these variances to isolate the component in \ocen from
 astrophysical effects.
We note that this approach to estimating
 the effect requires no assumption of the underlying distribution for the
 residuals from any one source.

Based on the evidence summarized previously, it is not unreasonable to assume that Reticulum represents
 a system for which the astrophysical sources of scatter 
 (metallicity, evolution, and period shift) are minimal. 
For Reticulum, the total variance, $\sigma_{tot}^{2}$, is the quadrature sum of the variance
 due to measurement uncertainties ($\sigma_{meas}$) and the intrinsic width of the PL ($\sigma_{int}$).
Thus, we can derive an upper limit on the intrinsic variance as,
\begin{equation}
 \sigma_{int} < \sqrt{\sigma_{tot}^{2} - \sigma_{meas}^{2}},
\end{equation}
 from which we estimate the intrinsic width of the PL is $\sigma_{int} < 0.03$ mag. 

The total variance \ocen must include an additional term for astrophysical effects ($\sigma_{astro}$).
We can similarly determine an upper limit on the added variance due to the combination
 of these astrophysical effects, as 
 \begin{equation}
 \sigma_{astro} < \sqrt{\sigma_{tot}^{2} - \sigma_{meas}^{2} -\sigma_{int}^{2}},
\end{equation}
from which we estimate the additional scatter in \ocen from astrophysical effects is 
 $\sigma_{astro} < 0.06$ mag, using our estimate of $\sigma_{int}$ from Reticulum.
 
In application to our Local Group RRL sample, we proceed
 to adopt an uncertainty due to astrophysical effects of $\sigma_{astro} < 0.06$ mag
 in the $H$ based on our comparisons to \ocen.
We will probe this effect directly in the $F160W$ magnitudes by observing a subset
 of \ocen RRL and combining these single phase observations with 
 our FourStar $H$ templates to derive mean magnitudes (see Table \ref{tab:rrltargets}).
A more in depth look at these effects requires a larger sample of 
 individual spectroscopic metallicities common to our NIR dataset 
 and one that fully samples the multiple stellar populations in \ocen. 

\begin{table*} 
\begin{center}
\caption{\label{tab:rrltargets} Summary of New Observations obtained for the RRL stage of the \cchp}
\begin{tabular}{l c c c c c l} 
 \hline \hline
 Target & Class & $\langle[Fe/H]\rangle$$^{1}$ & Supplementary Imaging & $F160W~(N_{obs})$ & $N_{RRL}$ & RR~Lyrae Discovery\\
 \hline \hline 
 \multicolumn{3}{l}{{\bf RR Lyrae Zero Point ---}} \\
    RR Lyr & Galactic RRab & -1.41$\pm$ 0.13 & TMMT & 2 Random  & & \\
    RZ Cep & Galactic RRc  & -1.77$\pm$ 0.20 & TMMT & 2 Random  & & \\
    UV Oct & Galactic RRab & -1.47$\pm$ 0.11 & TMMT & 2 Random  & & \\
    XZ Cyg & Galactic RRab & -1.44$\pm$ 0.20 & TMMT & 2 Random  & & \\
    SU Dra & Galactic RRab & -1.80$\pm$ 0.20 & TMMT & 2 Random  & & \\
 \hline
 \multicolumn{3}{l}{{\bf Metallicity Effects ---}} \\
    \ocen & Cluster    &  -2.5 to -1.1 & Magellan$+$FourStar & 2 Random  & 34 & \citet{sollima_2008}$^{3}$   \\
 \hline
 \multicolumn{3}{l}{{\bf TRGB Zero Point ---}} \\ 
    Sculptor   & dSph       & -1.68$^{2}$ & Magellan$+$FourStar   & 2  Random  &  52 & \citet{kaluzny_1995}$^{4}$ \\
    Fornax     & dSph       & -0.99$^{2}$ & Magellan$+$FourStar   & 2  Random  & 105 & \citet{bersier_2002}       \\
               &            &             &                       &            &     & \citet{mackey_2003}        \\
    IC\,1613   & IB(s)m     & -1.60$^{2}$ &                       & 12 Phased  &  67 & \citet{bernard_2010}       \\
    M\,32      & compact E2 & -0.44       &                       & 12 Phased  & 305 & \citet{fiorentino_2012}    \\
    M\,31      & SAB(s)b    & -1.50       &                       & 12 Phased  & 256 & \citet{sarajedini_2009}    \\
    M\,33      & SA(s)cd    & -1.48       &                       & 12 Phased  &  69 & \citet{sarajedini_2006}    \\
 \hline \hline
 \multicolumn{7}{l}{$^{1}$ Galactic RRL metallicities are from \citet{benedict_2011}
   and Local Group metallicities from Column 7, unless noted.} \\
 \multicolumn{7}{l}{$^{2}$ Mean stellar metallicity for RGB stars with medium resolution 
    spectroscopy for Sculptor and Fornax and from CMD modelling for IC\,1613 in the compilation of \citet{mcconnachie_2012}.} \\
 \multicolumn{7}{l}{$^{3}$ The most recent compilation of \ocen RR Lyrae is \citet{navarrete_2015}.} \\
 \multicolumn{7}{l}{$^{4}$ The most recent compilation of Sculptor RR Lyrae is \citet{martinezvazquez_2015}.}    \\
\end{tabular}
\end{center}
\end{table*} 

\subsection{Current Error Budget for the RR Lyrae} \label{ssec:rrlerror}

Assuming the intrinsic dispersion for a multi-abundance population is best matched to that of the total scatter in \ocen, we adopt $\sigma_{astro} = 0.06$ mag from \ocen for the intrinsic scatter term in our preliminary uncertainty estimate for the Local Group RRL pointings.
This term will be evaluated on a case-by-case basis for each of our RRL pointings in the \ho~ error budget at the conclusion of our program.
This is then coupled with the uncertainty of our zero-point discussion for $H$ ($\sigma_{fit}=0.05$ mag)
and we project a total uncertainty on the determination of a given distance using the PL of $\sigma_{total} = 0.079$ mag.
Converting to the error on the mean ($\sigma_{total}/\sqrt(5)$), we find an
  uncertainty on the mean of 0.036 mag, or 1.7 per cent in distance.
Adopting instead the scatter from a mono-abundance population from Reticulum of $\sigma_{H} = 0.04$~mag,
 we project an uncertainty on the mean of 0.030 or 1.4 per cent in distance.
At this stage in our program, in advance of the anticipated results from \gaia, 
 we select the conservative 1.7 per cent zero-point uncertainty for the RR~Lyrae $F160W$ PL zero point
 for our preliminary error budget. 

\section{The Absolute Scale for the TRGB} \label{sec:trgbabs}

The next step in the \cchp~ distance ladder is to determine the absolute magnitude of the TRGB.
The red giant branch terminates with a sharp discontinuity in its luminosity function
  that, when calibrated, can be used as a standard candle. 
This truncation of the RGB luminosity function is the result of the
 sudden lifting of degeneracy and rapid onset of helium burning
 throughout the isothermal core of the most 
 luminous RGB stars that represents the end of the RGB evolutionary phase \citep[][]{salaris_1997}.   
$I$-band observations of the TRGB have empirically shown this feature 
  to be very well delineated \citep[e.g.,][among others]{lee_1993,sakai_2004,caldwell_2006,rizzi_2007,mager_2008}.
 The TRGB can be determined by either  
   applying edge-detection techniques to the luminosity function on the RGB \citep[i.e.,][]{lee_1993}
   or fitting a modeled luminosity function to the observed luminosity function,
    when AGB and RGB populations cannot be separated \citep[i.e.,][]{mendez_2002}.
 The detailed study of \citet{rizzi_2007} furthered the technique as 
  a high precision distance indicator,
  where the precision is ultimately related to the number of stars near the TRGB
  \citep[as demonstrated explicitly in][]{madore_1995, madore_2009}.

A number of calibration issues remain to be addressed to take the TRGB method
 beyond its current precision to that of utility for our program
 \citep[the 1-$\sigma$ dispersion of the $I$-band 
 magnitude at the tip is estimated at $\sigma_{M_{I}} \pm$ 0.12~mag;][]{bellazzini_2008}; 
 the \cchp~ was explicitly designed to undertake a systematic study of these issues.
 The TRGB has an expected, but modest, dependence on metallicity; 
   however, an exact determination of the absolute magnitude of the TRGB 
   has remained challenging within the Milky Way. 
 Galactic globular clusters were the first objects used 
  to determine the zero point \citep[beginning with][]{dacosta_1990}, 
  but at the level of accuracy we are now seeking, there are many issues to address: 
  the distances to the globular cluster hosts themselves are uncertain due 
  to main sequence fitting uncertainties and line-of-sight reddening corrections, 
  on top of the systematic uncertainties in adopting a (multi-abundance dependent) fiducial main sequence. 
 Of greater concern is the fact that there is no guarantee for any 
  given globular cluster that there will be enough giants 
  currently ascending the RGB to fully populate the brightest part of
  luminosity function, thus leading to a systematic underestimate of the
  luminosity of the TRGB.
 For the \cchp,
  we will determine the absolute magnitude of the TRGB relative
  to other distance indicators in Local Group galaxies, in particular the RRL.

We now discuss our plans to set the absolute zero point of the TRGB
 method using two independent paths: 
(a) Using the geometric distance to the circum-nuclear maser disk 
    in the active galaxy NGC~4258 (Section \ref{ssec:path1}), and 
(b) Applying the parallax-based Galactic RRL near-infrared PL relation 
    to the halo populations of six Local Group galaxies (Section \ref{ssec:path2}). 
The error budget from this stage of the Population II distance ladder is given in Section \ref{ssec:trgberr}.
In the future with \gaia, we can by-pass the RRL distance scale entirely
 and calibrate the TRGB directly (Section \ref{trgbonly}).

\subsection{Path 1 -- The Geometric Distances to the Megamaser Host NGC~4258} \label{ssec:path1}
 
Currently, the only technique capable of measuring geometric distances
  outside of the Local Group is the utilization of H$_{2}$O 
 megamasers found in the accretion disks around
  central black holes. 
 If seen approximately edge-on and tracked over time,
  the individual maser sources can be used to trace the dynamics
  within the accretion disk and, under the assumption of Keplerian motion,
  constrain the line-of-sight distance to the host galaxy using geometry. 
 
 NGC\,4258 is the only galaxy near enough to be used as a TRGB calibrator. 
 It  has a geometric distance
  derived from the kinematics of its circum-nuclear maser disk \citep{humphreys_2013}.
 Modeling of the accretion disk using the detailed kinematical data 
  produces a line-of-sight distance of $d_{NGC\,4258} =$ 7.60 $\pm$0.17 (random) $\pm$0.15 (systematic)
  Mpc or $\mu_{NGC\,4258} = $29.40 $\pm$ 0.05 (random) $\pm$ 0.04 (systematic) mag.
  
\begin{figure*} 
\centering
 \includegraphics[width=\textwidth]{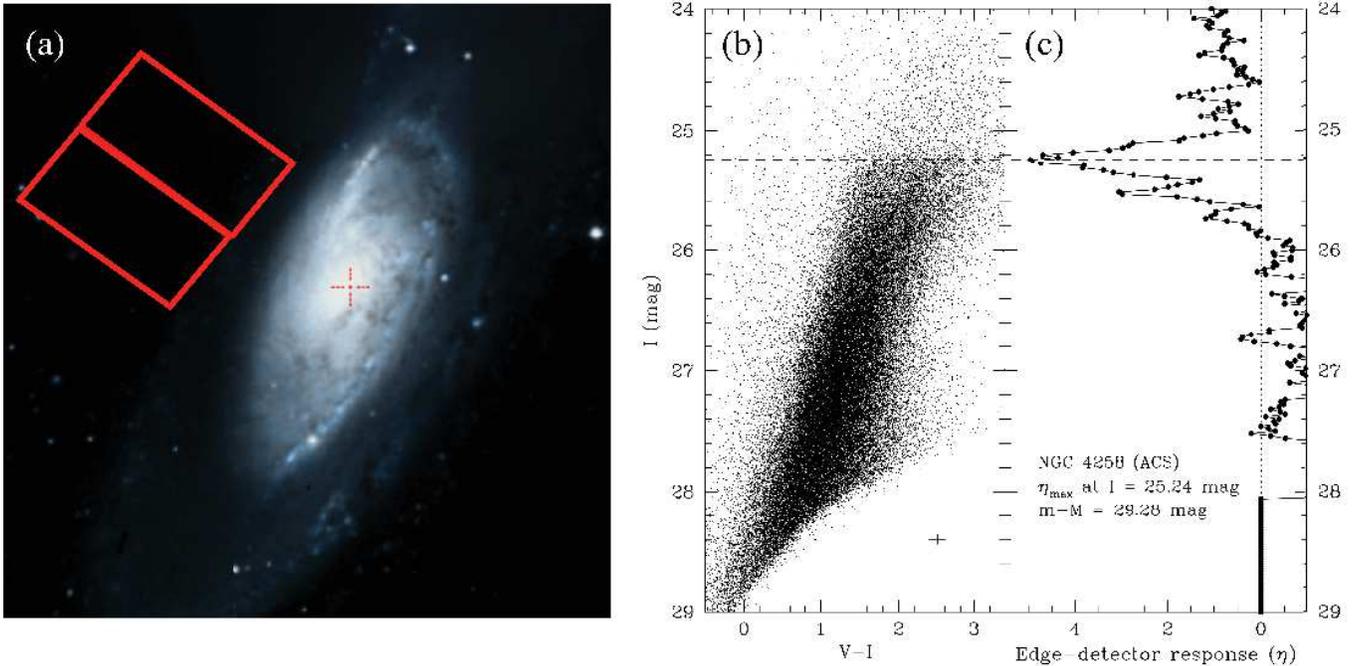}
\caption{Example TRGB detection in NGC\,4258.
  {\bf (a)} The location of the HST+ACS pointing of 
   \citet{mager_2008} (blue outline) relative to the disk of NGC\,4258 
    (from DSS) used to determine the TRGB distance to NGC\,4258,
   emphasizing its placement securely in the halo of NGC\,4258. 
  {\bf (b)} The $HST+ACS$ $V$,$I$ CMD in the halo of NGC\,4258 reveals
   a high contrast TRGB detectable in a galaxy with $\mu$\textgreater29.0~mag. 
  {\bf (c)} The corresponding response of an edge detection filter,
   effectively the first derivative of the $I$ luminosity function,
   applied to the CMD shows a large peak consistent with the visual 
   impression of the TRGB (dotted line).
  The $I$ magnitude has been adjusted for metallicity effects. 
   Reproduced from \citet{mager_2008}.
   \label{fig:ngc4258cmd}}
\end{figure*}

 \citet{mager_2008} measured TRGB stars in the halo of this same galaxy 
  using $HST+ACS$ observations \citep[PID 9477]{hst_2002},
  specifically designed to target its Population II rich stellar halo.
 Figure \ref{fig:ngc4258cmd}a demonstrates the location of the $HST+ACS$ field
  and Figure \ref{fig:ngc4258cmd}b presents the exquisite red giant branch
  with a highly resolved TRGB.
 We note that the metallicity dependence of the TRGB, 
  which is projected into a color dependence in the CMD, 
  has been calibrated out
  of Figure \ref{fig:ngc4258cmd}b following the procedure discussed in \citet{mager_2008}
  and more formally presented in \citet{madore_2009}.
 The results of application of an edge-detection algorithm to the luminosity function
  is shown Figure \ref{fig:ngc4258cmd}c and 
  determines the TRGB at $I$ = 25.24 $\pm$ 0.04 \citep[][]{mager_2008}.
 
 Combining the maser distance with the TRGB estimate, we obtain
  an absolute magnitude for the $M_{I}^{TRGB}$= -4.16 $\pm$ 0.06 (random) $\pm$ 0.04 (systematic) mag.
 This value is broadly consistent with other calibrations of the TRGB, 
  $M_{I}^{TRGB} = -4.04\pm$ 0.12 mag for [Fe/H]$=-1.52$ \citep{bellazzini_2001,bellazzini_2004},
  which is the relation used by \citet{mager_2008} to calibrate the dependence of the TRGB on metallicity, 
  but is slightly offset.
 We note that the initial maser distance at the time of the \citet{mager_2008} from \citet{herrnstein_1999}
  was entirely consistent with both the TRGB and Cepheid based distances.

 Within the context of the \cchp, we use these existing observations to provide a 
  direct calibration of the TRGB, beyond those shown in Table \ref{tab:trgbtargets}.
 More specifically, we are re-analyzing these archival data to ensure
  identical image processing techniques and photometry are employed.
 We plan to calibrate the TRGB in the native $F814W$ \hst-\emph{ACS} flight
  magnitude system to eliminate uncertainties imposed by filter conversions.

\subsection{Path 2 -- The Local Group RR~Lyrae Variables} \label{ssec:path2}

 Using the RR~Lyrae PL discussed in the previous section, 
  we can determine precise distances to Local Group galaxies.  
 Each of these Local Group galaxies has a stellar halo rich in resolved
  Population II RGB stars
  and we, thereby, can place the TRGB absolute magnitude firmly onto the RRL geometric distance scale.
 
 We have selected galaxies in the Local Group that have known and well-characterized RRL 
   (i.e., with light curves sufficient to identify the pulsation mode and to determine the period).
 To avoid effects from multiple populations, line-of-sight depth, and differential extinction, among others,
  we explicitly opt to exclude the Large Magellanic Cloud 
  or the inner regions of other nearby star forming galaxies not seen face-on. 
 Imposing these requirements yields six nearby galaxies:
  M\,31, M\,32, IC\,1613, M\,33, Sculptor, and Fornax,
  that have the prerequisite discovery data with \emph{HST}+ACS published in the literature)
   (specifics are given in Table \ref{tab:rrltargets}). 
 Due to the galactic demographics of the Local Group, five of our six calibrators 
  are dwarf galaxies, but given that these systems are the primary building blocks of the 
  galactic halos we will probe in the \sn hosts, 
  the stellar content of these dwarf galaxies is likely representative of these
  larger galaxies.

 \begin{table*} 
 \begin{center}
 \caption{\label{tab:trgbtargets} Summary of New Observations obtained for the TRGB stage of the \cchp }
\begin{tabular}{c c c c c l} 
 \hline \hline
  Target & Class$^{1}$ &  Supplementary Imaging & $R_{gal}$ & Cepheid Distance? & Notes \\
 \hline \hline 
  \multicolumn{3}{l}{{\bf TRGB Zero Point}}    \\
    Sculptor   & dSph        & Magellan+IMACS   &  3.3$'$ (0.08 kpc) & N &   \\
    Fornax     & dSph        & Magellan+IMACS   &  6.2$'$ (0.25 kpc) & N &   \\
    IC\,1613   & IB(s)m      & Magellan+IMACS   & 10.2$'$ (2.17 kpc) & Y &   \\
    M\,32      & compact E2  &                  &  2.9$'$ (0.64 kpc) & N &   \\
    M\,31      & SAB(s)b     &                  & 20.4$'$ (4.62 kpc) & Y &   \\
    M\,33      & dSpr        &                  & 15.9$'$ (4.08 kpc) & Y &   \\
  \hline
  \multicolumn{3}{l}{{\bf \sn Zero Point}} \\
    M\,101     & SAB(rs)cd        &  & 11.6$'$ (23.1 kpc) & Y & \sn: 2011fe  \\
    M\,66      & SAB(S)b          &  &  4.0$'$ (11.5 kpc) & Y & \sn: 1989B   \\
    M\,96      & SAB(rs)ab        &  &  4.2$'$ (13.2 kpc) & Y & \sn: 1998bu  \\
    NGC\,4536  & SAB(rs)bc        &  &  3.0$'$ (13.4 kpc) & Y & \sn: 1981B   \\
    NGC\,4526  & SAB0(s),edge-on  &  &  3.2$'$ (14.2 kpc) & N & \sn: 1994D   \\
    NGC\,4424  & SAB(s)           &  &  2.3$'$ (10.3 kpc) & N & \sn: 2012cg  \\
    NGC\,1448  & SAcd, edge-on    &  &  2.6$'$ (13.3 kpc) & N & \sn: 2001el  \\
    NGC\,1365  & SB(s)b           &  &  3.6$'$ (19.1 kpc) & Y & \sn: 2012fr  \\ 
    NGC\,1316  & SAB(0)  pec      &  &  9.2$'$ (53.6 kpc) & N & \sn: 1980D, 1981D, \\
               &                  &  &                    &   & 2006dd, 2006mr  \\
   \hline \hline
\multicolumn{5}{l}{$^{1}$ From the NASA Extragalactic Database.}
\end{tabular}
\end{center}
\end{table*} 

\subsubsection{New HST Observations} \label{sssec:trgbnewobs}

\begin{figure*}  
 \centering
 \includegraphics[width=\textwidth]{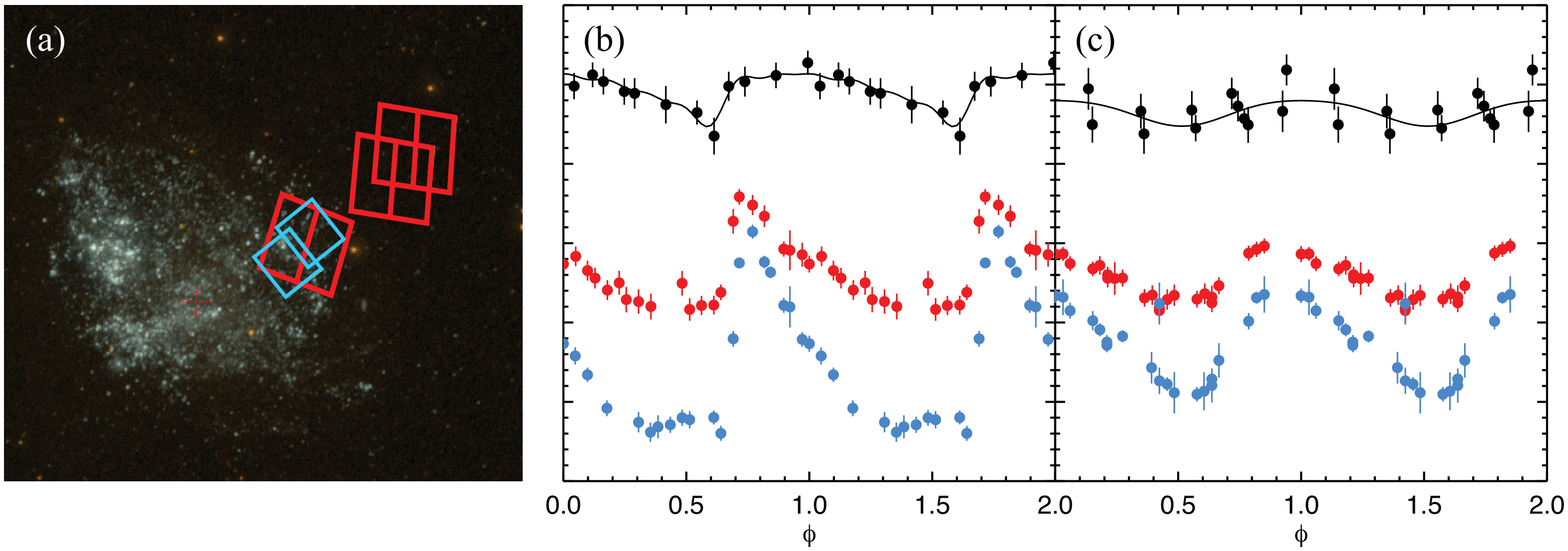}
 \caption{Example Local Group RR Lyrae and TRGB data for IC\,1613.
  {\bf (a)} Image demonstrating our \hst$+WFC3-IR$ RR Lyrae 
   and \hst$+ACS$ TRGB observations from the LCID project (inner fields)
   and our ACS parallel fields in the inner halo of IC\,1613 (outer fields).
  The background image is from \emph{GALEX}. 
 Example light curves for {\bf (b)} a type RRab (V039) with period 0.520 days
  and for {\bf (c)} a type RRc (V130) with period 0.314 days {\bf (Hatt et al.~in prep.)}. 
 Sample \hst light curves in the optical 
  \citep[$F475W$, blue, bottom; $F814W$, red, middle][]{bernard_2010}, 
  and near-infrared ($F160W$, black, top; Hatt et al.~in prep.).
 {\bf As a part of this program, we derive new photometry for the literature data,
  but adopt the period measurements from earlier discovery work.}
 A black line shows the best-fit \hst near-infrared 
  imaging template via \citep[][]{yang2012}.  
 \label{fig:hbandlightcurve}}
\end{figure*}

As discussed in Section \ref{ssec:rrlerror}, 
  we have used the $HST+WFC3-IR$ channel with the $F160W$ filter for our studies of the RR~Lyrae. 
 As in the CHP-I for Cepheids \citep[for example light curves see][]{monson_2012, scowcroft_2011}, 
  we obtain 12 phase points 
  that are used with our template fitting techniques (Section \ref{ssec:templates})
  to determine the phase-weighted mean magnitudes.
 This combination of these two techniques can produce mean magnitudes at the 2 per cent level for individual stars
  \citep[see detailed discussions in][]{madore_2005,scowcroft_2011}. 
 Example data from our pointing in IC\,1613 is given in Figure \ref{fig:hbandlightcurve}, 
  where Figure \ref{fig:hbandlightcurve}a shows the pointing map 
  (the \hst$+WFC3-IR$ pointing is smaller FOV) 
  and example light curves in $F475W$ (blue), $F814W$ (red), and $F160W$ (black) 
  for RRab and RRc variables in IC~1613 are shown in panels \ref{fig:hbandlightcurve}b
  and \ref{fig:hbandlightcurve}c, respectively.
The data presented in Figures \ref{fig:hbandlightcurve}b and \ref{fig:hbandlightcurve}c 
   will be presented in full by in Hatt et al \emph{(in prep)} and includes
   a comparison to both published works using other methods, as well as our
   first estimate of the $F160W$ TRGB zero point.
 Figure \ref{fig:hbandlightcurve} demonstrates that we have sufficient light curve sampling
  to determine mean magnitudes to high precision. 
  
 The RRL observations for each of the six galaxies are effectively identical,
  with the exception of Fornax and Sculptor (to be discussed below),
  and are summarized in Table \ref{tab:rrltargets}. 
 Fields were chosen to have about 100 RR Lyrae 
  that were previously characterized in at least two optical photometric bands.
 We anticipate some fraction of the objects in each field will be blended or crowded due both 
  to the change in resolution from 
  their detection in \hst$+ACS$ at 0.05$''$ pixel$^{-1}$ to \hst$+WFC3-IR$ at 0.135$''$ pixel$^{-1}$
  and to the increase in the relative brightness of the AGB populations at this wavelength.
 To cover the majority of the original \hst$+ACS$ field, 
  two \hst$+WFC3-IR$ fields are required per single \hst$+ACS$ field and we optimized the placement of these fields
  to maximize the total number of RR Lyrae. 
 We also intentionally placed some overlap between the two fields to test for systematic effects.
  
 The RRL observations for \ocen, Fornax and Sculptor are designed to obtain closely spaced phase points 
  during a single \hst orbit in $F160W$.
 For both Fornax and Sculptor, 
  parallel ACS observations for the purpose of the TRGB are also obtained
  as to be discussed in Section \ref{ssec:grdsupplement}
  (exposures in \ocen were too frequent to permit parallel observations due to data transmission rates).
The RRL observations will be combined with literature optical light curves (see Figure \ref{tab:rrltargets})
 and the self-template technique of Section \ref{ssec:templates} is applied to determine mean magnitudes.

 For the TRGB observations, the Local Group targets can be split into two categories:
  those of low surface brightness and high surface brightness. 
 For the low surface brightness galaxies, a single ACS field does not 
  have sufficient sampling of the TRGB and
  the photometry for these objects has been supplemented with ground based observations (Section \ref{ssec:grdsupplement}).
 For the high surface brightness galaxies, the ACS field does has sufficient 
  statistics to measure the TRGB directly, as we now describe.
 The TRGB data for the Local Group objects are obtained with ACS in parallel mode to the
  RRL data previously described.
  
 We constrained the roll angle of the HST observations such that the ACS field  occurs
  at larger object-centric radii, both safely within their halos
  and at sufficiently low stellar density to avoid significant crowding.
 For clarity, we include an example Local Group pointing in Figure \ref{fig:hbandlightcurve}a,
  for IC\,1613 to demonstrate the coupled observations
  and their relationship to the archival \hst imaging.
 We are also using imaging data from the archived HST RRL discovery programs (as noted in Table \ref{tab:rrltargets})
  to measure directly the TRGB magnitude in these fields to improve statistics
  and check for slight differences, if any, in crowding and metallicity between the fields.
  
\subsubsection{Supplementary Wide Field Observations} \label{ssec:grdsupplement}

 Sculptor, Fornax, and IC\,1613 are each dwarf galaxies with intrinsically low stellar density. 
 While wide-field observations that contain the full extent of the galaxy (to several half-light radii)
  sufficiently sample the TRGB, 
  our relatively small ACS field ($4' \times 4'$) samples a relatively small area of these nearby systems
  ($\sim$4 per cent for Sculptor, $\sim$2 per cent for Fornax using the half-light radius given in \citet{mcconnachie_2012};
   $\sim$8 per cent for IC\,1613 using the profile fits of \citet{battinelli_2007} to the old populations of the ``spheroid''). 
 Statistical studies of TRGB accuracy conclude that approximately 100 stars within one magnitude of the TRGB 
  are necessary to reach the precision necessary for our program \citep{madore_1995,madore_2009}.
 Thus, we supplement our \hst data for these objects with a ground based, wide-field campaign using 
  Magellan+IMACS, which, with a field-of-view \textgreater $15' \times 15'$, 
  provides a larger area by a factor of 14 for a single pointing. 
 By doing multiple pointings, we can sample the full half-light area for these three objects.
  
 The ground data are `bootstrapped' from the ground based Johnson cousins (i.e., $V$ and $I$) system
  to the native flight magnitude system of HST (i.e., $F814W$ and $F606W$).
 All stars common to the two datasets can be employed for the bootstrapping to ensure
  appropriate precision is obtained
  and to investigate for colour-terms in the transformation. 
 The magnitude of the TRGB for these galaxies is then determined on the photometry of this 
  more populated dataset calibrated to $F814W$. 
 
\subsection{The Current Error Budget for the TRGB Calibration} \label{ssec:trgberr}

 The TRGB method has been widely applied and extensively tested by many groups. 
 Most recently, \citet{rizzi_2007} reduced and analyzed in a
  uniform manner HST $V$ and $I$ data. 
 \citet{rizzi_2007} find that the mean statistical uncertainty in an individual 
  distance estimate using the TRGB is only 0.05~mag or 2.5 per cent in distance. 
 The geometric distance to the maser host galaxy NGC~4258 has also been measured
  to an accuracy of 2 per cent in distance
  and its TRGB magnitude has also been measured to 2 per cent. 
 Our new seven independent (Local Group and NGC\,4258) 
  estimates of the TRGB absolute magnitude will result in an error 
  on the mean of 0.05/$\sqrt{7}$ = 0.019~mag. 
 For the purposes of understanding our \cchp~ error budget,
  we project an error of 0.9 per cent for this stage of the Population II distance ladder.

\section{TRGB Distances to \sn Hosts} \label{sec:trgbsn1a}

The next step in our Population II distance ladder is to calibrate
 the absolute magnitude for \sn.
This is accomplished by applying our TRGB zero point (Section \ref{sec:trgbabs})
 to TRGB measurements in \sn host galaxies, as we now describe.
We discuss the use of \sn for distance measures (Section \ref{ssec:sndistance}),
 observational concerns over the use of the TRGB (Section \ref{ssec:pushingtrgb}),
 and estimate our error budget for this step (Section \ref{ssec:snzperror}).

\subsection{\sn as Distance Indicators} \label{ssec:sndistance}
\sn are the {\it de facto} standard candles for
 determining distances at cosmologically significant scales.  
\sn, however, are not trivial objects to characterize.
Once discovered in transient surveys or by serendipity, 
 the transient event must be typed with spectroscopy,
 followed with time series, multi-band photometry for up to $\sim$ 30 days 
 to identify maximum light from which the proper decline rate correction is determined,
 and a long-baseline photometry is required to determine a precise
 host galaxy template from which the internal component of reddening can be estimated 
 \citep[see][for a good description of the required data for \sn characterization]{hamuy_2006}.
The need for such studies to be undertaken with as few systematics as possible
 has inspired several large scale \sn surveys, including
 the \emph{Carnegie Supernova Project} \citep[CSP, hereafter; see][for descriptions of the process]{freedman_2009,folatelli_2010,burns_2014}.
 
The internal scatter (i.e., the precision) of \sn luminosities 
 is found to be $\sim0.16$ mag \citep[e.g.,][]{folatelli_2010}.
\sn are intrinsically rare events and it is even more unusual
  to have one occur sufficiently close to the Milky Way that the
  distance to its host galaxy can be independently measured. 
In the last quarter century there have been only 19 \sn found 
  and measured (using modern, linear detectors) in spiral galaxies 
  that are close enough for their distances 
  to be determined independently by Cepheids.\footnote{We note that 11 of these 19 galaxies were presented in \citet{riess_2016} for the first time.}

Beyond the limited sample size, there are other concerns with
 respect to the \sn. 
In particular, there is evidence \citep[e.g.,][among others]{sullivan_2010,rigault_2015} that the
  absolute magnitudes of \sn depend on whether they are hosted by
  lower--mass (spiral and/or irregular) or high--mass (elliptical and/or
  lenticular) galaxies. 
The Cepheid distance scale is only applicable to \sn whose hosts 
 contain significant young (20-400 Myr) populations,
 predominantly spiral galaxies.
Thus, a Cepheid based calibration of the \sn absolute magnitude is subject to any 
 population effects within the \sn sample
 (regardless of whether systematic or random in nature). 
Our Population~II route will immediately bring \sn hosted by non-spiral systems 
  into the calibration sample, 
  and provides a direct test of such population differences.
Additionally, the use of Population~II distance indicators, 
 permits use of \sn in host galaxies where the 
 Cepheid population is not suitable due to crowding or large extinction,
 as occurs commonly for edge-on galaxies.

Regardless of Hubble type, galaxy luminosity, and inclination, 
 hundreds of TRGB stars populate the dust-free, 
 low-metallicity, low-crowding stellar halos.
Nine \sn have been found in such galaxies in the past quarter century; 
 all but one of them have good quality (non-photographic) light curves. 
The TRGB is not without pitfalls, the most important of which
 is the need to choose pointings appropriate for the method
 to reliably produce measurements of the quality required here. 
 
\subsection{Choice of Fields for the TRGB Method} \label{ssec:pushingtrgb}

Before discussing the calibration of the TRGB further, 
 there are several challenges of the TRGB as a distance indicator that are
 worth consideration in the development of a Population II distance scale. 
Two important examples of applying the TRGB method inappropriately are
 M\,101 and NGC\,4038/39. 
As we will discuss in application to these two case studies, 
 there are clear reasons, either in the observational data 
 or in the edge-detection itself, that drive the dispersion in published distance estimates. 

There are four published TRGB distance estimates to M\,101 in the literature: 
 they range from $29.05\pm0.06$ (random)$\pm 0.12$ (systematic)~mag \citep{shappee_2011}, 
  to $29.42\pm0.04$ (random)$\pm 0.10$ (systematic)~mag \citet{sakai_2004}, 
  to $29.34\pm0.08$ (random)$\pm 0.02$ (systematic)~mag \citet{rizzi_2007}, 
  and then most recently a value of 
  $29.30\pm0.01$ (random)$\pm 0.12$ (systematic)~mag \citet{lee_2012}
   based on the weighted mean of nine individual fields.
These studies present a large range of moduli, spanning almost 0.4~mag
 \citep[in contrast, the Cepheid based distances span a range of 0.7 mag, see compilation in][]{lee_2012}.  
Not only do the cited distances span a large range, but this sample demonstrates the broad range
 of uncertainties quoted (both random and systematic) by various studies.

As it stands this is an unacceptable outcome 
 for ultimately calibrating the Type~Ia supernova.
Unfortunately, the fields used to determine the TRGB distances were based on the same data used to
 study the Cepheids.  
Owing to their initial purpose, 
 all of those fields are well into the disk of M\,101
 and are therefore very crowded at the significantly fainter magnitude level of the TRGB.
Moreover, they are heavily contaminated by brighter, intermediate-aged, asymptotic giant
  branch (AGB) and post-AGB stars, 
  that confuse and bias any attempt(s) to identify a pure sample of TRGB stars in the optical bands. 

This latter problem and its negative consequences are not new. 
\citet{saviane_2004, saviane_2008},
 using data deep into the star-forming regions of NGC\,4038/39
 and contaminated by a brighter and dominant population of AGB stars,
 measured the distance moduli to be $\mu = 30.7\pm0.25$ (random) $\pm0.14$ (systematic)~mag and 
  $\mu = 30.62\pm0.09$ (random) $\pm0.14$ (systematic)~mag, respectively. 
This was soon rectified by \citet{schweizer_2008},
  who measured the TRGB at 
   $\mu = 31.51\pm0.12$ (random) $\pm 0.12$ (systematic)~mag. 
This measurement pushed the host galaxy out to 20.0~Mpc,
  nearly a 50 per cent increase over the distance of 13~Mpc suggested by the initial studies. 
More recently, \citet{jang_2015} have independently confirmed the result of \citet{schweizer_2008}.
Again, high quality distances can be determined if proper care is taken to either avoid the high stellar density disk 
 or to systematically exclude crowded regions from the TRGB fit.

Further confirmation of the importance of carefully selecting TRGB halo fields comes from
 the complementary studies of \citet{caldwell_2006} and \citet{durrell_2007},
 each of which used \hst+ACS to measure TRGB distances to individual Virgo cluster dwarf galaxies 
 and, in \citet{caldwell_2006}, to intra-cluster stellar populations.
These observations were designed with the TRGB in mind and targeted regions of low stellar density
 with limited contamination from young and intermediate-aged populations.
 A mean distance modulus of $\mu = 31.05 \pm 0.05$ mag or $d = 16.1 \pm 0.4$ Mpc 
 was obtained for the nine fields (seven dwarfs and two intermediate regions).
Individual distance moduli have quoted uncertainties of $\sim$0.10~mag and span a range of 0.3~mag. 
 We note, however, that identical methods and data collection strategies were used in these
  works and confirmation by an independent team (as was the case 
   for M\,101 and NGC\,4038/39) is required to verify these measurements.
Nevertheless, as was shown for NGC\,4258 in Figure \ref{fig:ngc4258cmd}
 and has been demonstrated in the literature for M\,66 and M\,96 \citep{lee_2013}, 
 M\,74 \citep{jang_2014}, and NGC\,5584 \citep{jang_2015}, in addition to
 those objects previously described,
 halos provide excellent environs for application of the TRGB
 as a distance indicator.
 
In the cases where the TRGB is measured using imaging of appropriate galactic locales,
 i.e. those in which the old, metal poor population dominate, 
 and in which careful consideration is given to how to measure the 
 discontinuity, 
 the TRGB proves to be of comparable precision as other distance indicators 
 at these faint magnitudes.

\begin{figure*}  
\centering
\includegraphics[width=0.70\linewidth]{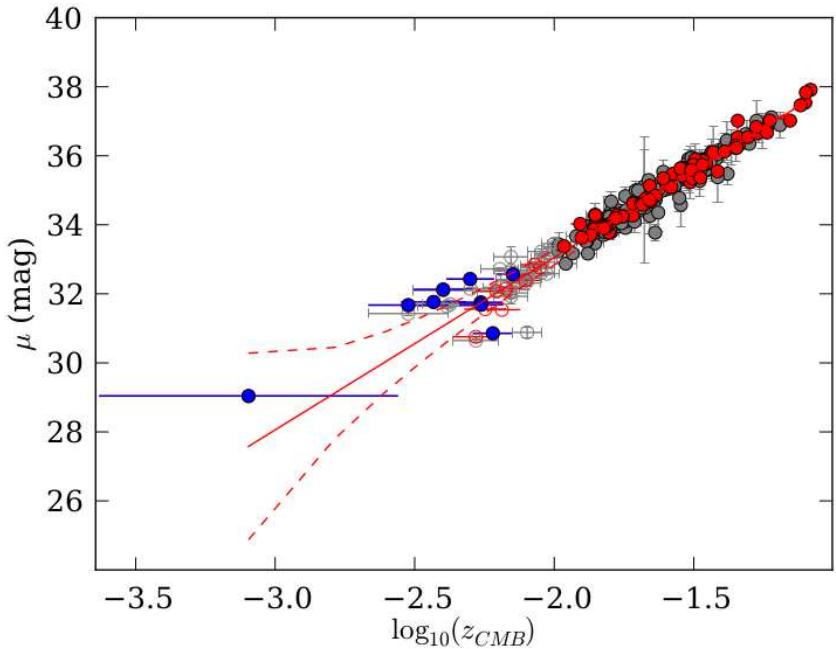}
 \caption{ \label{fig:hubbleflow} Illustrative \sn Hubble Diagram comparing the
   distance modulus and redshift for \sn. 
  Grey symbols are data from the literature \citep[primarily from the CfA4 data release,][]{hicken_2012}.
  Red symbols are those supernovae included in the CSP-1 \citep[e.g.,][]{stritzinger_2011}.
  Blue symbols are the local supernova calibrators, 
   demonstrating the larger scatter induced by local large-scale flows for nearby objects. 
   }
\end{figure*}  

\subsection{The \sn Zero Point Error Budget} \label{ssec:snzperror}

 In our sample, there are nine galaxies that have hosted twelve well-observed \sn~(Table \ref{tab:trgbtargets}),
  with the most distant object, NGC\,1316, hosting four \sn. 
Our observing strategy for these galaxies was based on the study of \citet{caldwell_2006},
  who measured the TRGB for the equally distant Virgo cluster dwarfs with ACS 
  to a precision of 0.1~mag. 
 However, we adopt the much higher throughput filter $F606W$, 
  as opposed to $F555W$ as used by \citet{caldwell_2006}, to increase observing efficiency. 
 The \hst observations (Table \ref{tab:trgbtargets}) 
  were designed to yield a 10$\sigma$ detection at the TRGB in the $F814W$ filter
  and, since the color will only be used to remove contaminants,  
  a $\sim$3$\sigma$ detection in $F606W$ for stars at the anticipated TRGB in each galaxy.
 Preliminary reductions of one of our more distant objects, NGC\,1365, 
  to be presented by Jang et al.~(in prep), 
  suggests this signal-to-noise goal is being met. 

Field choice for our \hst imaging was governed by the following general strategy.
First, we avoided extended disks and other obvious young or tidal structures. 
Second, we attempted to target preferentially along the minor axis to minimize 
 the likely contamination from low surface-brightness thick disk structure.
Lastly, we straddled the \emph{WISE} W1 25-26 mag arcsec$^{-2}$ isophote and, if applicable, 
 the \emph{GALEX} Near-UV 27-28 mag arcsec$^{-2}$ isophote (using custom builds of the public raw data).
This resulted in a sample that ranges in galacto-centric distances ranging from
 $\sim$ 10 kpc in NGC\,4424 (one of the least luminous galaxies) 
 and $\sim$ 60 kpc in NGC\,1316 (having prominent shell features and being one of the most massive galaxies).
Total anticipated RGB source density was confirmed using 
 published data from \citet{mager_2008} for NGC\,4258 and
 from \citet{lee_2013} for M\,66 and M\,96, 
 as well as providing verification of our signal-to-noise estimates.

Using the metallicity profile of \citet{gilbert_2014} for the resolved stellar halo
 of M\,31 (between $R_{proj}$ $\sim$10 kpc and 165 kpc) as a proxy for our galaxies,
 we can expect a M\,31-analoge galaxy to have median stellar metallicities between $\sim$-0.5 dex
 and $\sim$-1.0 dex over this radius range \citep[see Figure 11 of][]{gilbert_2014}.
The dispersion of the median metallicity is $\sim$0.5 dex at any given location \citep[see Figure 9 of][]{gilbert_2014}.
Thus, the range of metallicities we expect in our \sn hosts is representative of the 
 range of [Fe/H] spanned by our Local Group and Galactic targets (Table \ref{tab:rrltargets}).
Thus, our TRGB ZP determination relative to the RRL will include variations over this metallicity range.

A great benefit of the metallicity dependence of the TRGB is that, unlike the dependence for RRL and Cepheids,
 the effect of metallicity is encoded into the color of the RGB star,
 and the most metal poor stars -- those that dominate in our haloes -- 
 are brighter than the metal rich stars.
 Thus, adding additional metal rich stellar populations will `blur'
  the tip detection preferentially to fainter magnitudes. 
 The total impact on our TRGB measurements will depend on the star formation history 
  of a given pointing and we can use our M\,31, M\,33 and M\,32 TRGB data to study these
  effects empirically within the context of our program.
 For the purposes of this discussion, we assume an equal distribution of stars over our color range. 
 The impact of metallicity on the TRGB detection has been studied by 
 \citet[][among others]{rizzi_2007,mager_2008,madore_2009} 
 and these studies consistently find a TRGB slope of $\sim$0.2 mag color$^{-1}$ 
 for both the ground based Johnson filters and the \hst~ flight magnitude system. 
 Using the published TRGB results of \citet{lee_2013}, we can expect a well populated TRGB 
  over a 0.5 mag color range in $F606W-F814W$, which implies a uniform change
  in the TRGB magnitude of $\sim$ 0.1 mag in $F814W$ 
  \citep[using the TRGB slope in the flight magnitude system from][]{rizzi_2007}.
 Converting the uniform change to an effective dispersion, we predict a blurring over 
  this color range of $\sigma_{[Fe/H]}$ = 0.028 mag.

Taking considerations from the previous sections to heart, as part of our program, 
 we plan to reanalyze the \citet{lee_2013} pointings for M\,96 and M\,66 (GO-10433),
 as well as extensive archival fields (taken as parallel pointings) 
 available for NGC\,4258 and M\,101 that reach a similar signal-to-noise as our primary program.
 This will both provide independent checks on our methods (i.e., in comparison to previous work),
  as well as providing direct empirical tests for AGB-contamination, crowding,
  and metallicity effects as a function of galacto-centric radius for these objects.

For our final measurement of \ho~ from this distance ladder,
  individual fitting errors, derived independently for each galaxy and including 
   constraints on the effects of metallicity (color), crowding, and completeness, 
  will be used (a first exploration of this will be presented for NGC\,1365 by Jang et al. in prep).
 For the goals of this work, we make an estimate of the uncertainty for the \cchp~ error budget,
   by adopting twice the TRGB fitting uncertainty from 
   \citet{rizzi_2007} ($\sigma = 0.10$~mag)
   to account for increased errors in the
   $I$-band photometry at the tip for our more distant sample of \sn hosts,
   a value that is consistent with the uncertainties quoted by \citet{caldwell_2006} for the Virgo cluster.
We adopt a TRGB blurring of $\sigma_{[Fe/H]}$ = 0.028 mag, which when added in quadrature to
  our anticipated TRGB fitting uncertainty is negligible.
 To our total TRGB uncertainty, 
  we then add in quadrature the 0.12~mag intrinsic scatter in the \sn absolute magnitudes 
  \citep[with the term for large scale flows removed, see][]{folatelli_2010} 
 and obtain a projected uncertainty for our calibrating sample of $\pm$0.157~mag per source. 
 For the pure Population II calibration using twelve TRGB-calibrated 
  \sn we obtain a systematic error on the mean of 
   0.157/$\sqrt{12}$ = 0.046~mag or 2.1 per cent in distance. 
 Thus, we project a 2.1 per cent uncertainty for this stage of the distance ladder.

\section{Into the Hubble Flow} \label{sec:hubbleflow}

Given a larger sample of \sn calibrating galaxies, 
 augmented by \sn calibrated by Population~II distance indicators, 
 directly traceable to parallaxes, we will derive a
 new value of \ho based on several hundred \sn at distances well into the smooth Hubble flow. 
Well-observed light curves for these \sn can be directly adopted from 
 CSP1 \citep[][and references therein]{stritzinger_2011},
 CfA4 \citep[][and references therein]{hicken_2012}, 
 Pan-STARRS1 \citep[][and references therein]{rest_2014}, 
 and CfAIR2 \citep[][and references therein]{friedman_2015}.
Conservative selection of low-z \sn with low internal extinction from these data releases
 suggest approximately 221 suitable objects are available now.
By the completion of our program, we anticipate additional \sn light curves to be released, 
 both from additional data released from the more recent incarnations of the CSP and CfA programs
  and from the continued operations of Pan-STARRS
  ({\it The Foundation Survey}; R. Foley, priv. communication),
  the Dark Energy Survey (DES), and other large-area, multi-epoch imaging programs.

Figure \ref{fig:hubbleflow} is a Hubble diagram that uses data from 
 CfA4 \citep[][]{hicken_2012} and the
 the CSP1 \citep[][]{stritzinger_2011} to demonstrate this last step in our distance ladder
 (we note this is not the full sample available now in the literature). 
Figure \ref{fig:hubbleflow} contains 130 \sn from CSP (red and blue points)
  and 94 from CfA4 (grey). 
The \sn presented in Figure \ref{fig:hubbleflow} were selected to those that 
 had first epoch observations within 10 days post maximum 
 and with light curve stretch factors less than 0.5 
  (i.e., fast decliners are excluded) for a total of 215 objects with redshifts larger
  than 0.1 (64 objects were from CSP).
The CSP \sn were observed in the near-IR, 
 which has two benefits, decreased extinction
 and decreased decline rate corrections \citep[for details, see the review by][]{phillips_2012}, 
 that improve the resulting measurements
 (compare the variance of the red and grey points in Figure \ref{fig:hubbleflow}).
By the conclusion of our program, we anticipate Figure \ref{fig:hubbleflow} to have evolved 
 significantly with the continued operations of the \sn characterization programs previously mentioned.

This final step in the Population II distance scale is probably the most secure in its
 precision and accuracy given the small scatter of the \sn that are safely in the 
 Hubble flow \citep[$\sigma =0.16$ mag][]{folatelli_2010}.
At the same time, Figure \ref{fig:hubbleflow} also demonstrates the large scatter
 in the ``local'' \sn sample calibrated to Cepheids (blue points),
 whose line-of-sight motions have a higher fractional impact from their peculiar motions.
The variance shown in Figure \ref{fig:hubbleflow} will only improve
 with the on-going projects to better quantify the Hubble flow.

Using a representative sample (i.e., Figure \ref{fig:hubbleflow}),
 we can estimate the precision afforded by this stage of the \cchp~ distance ladder.
A dispersion of 0.16~mag for \sn around the Hubble line \citep{folatelli_2010} gives, 
  for a sample of 221 low-z and low-reddening \sn hosts, 
  a negligible error on the mean of only 0.16/$\sqrt{221}$ = 0.011~mag. 
We note that the number of \sn will only increase over the progression of the \cchp.
The dispersion estimate from \citet{folatelli_2010} already includes 
 the uncertainty arising from averaging over large-scale flows 
 \citep[$\sim$0.5 per cent;][following the prior analysis of Hicken et al.~1995]{riess_2009}.
At this stage of our program, we anticipate a final combined error of 0.5 per cent.
 
\begin{figure*}  
\includegraphics[width=\textwidth]{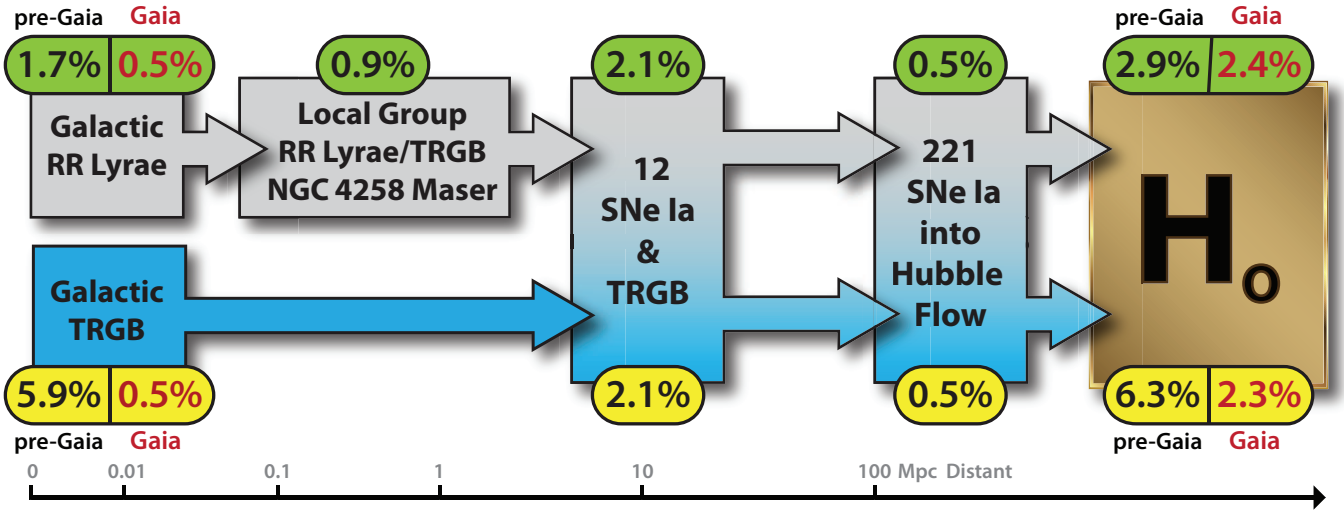}
\caption{\label{chpflowchart}
  Schematic representation of the projected error budget for \ho with only
   Population II distance indicators using the conservative estimates described in the text,
   where the current budget is given in black text and the projected budget after \gaia 
   is in red text. 
  Two paths are given; 
   the first is that described in the text that uses RR Lyrae to calibrate the TRGB and the TRGB to calibrate the \sn(top fork) 
 and the second is a path that calibrates the TRGB directly and then uses the TRGB to calibrate the \sn(bottom fork). 
  The scale bar at the bottom of the figure gives the approximate line-of-sight distance scale
  in units of Mpc for the measurements taken at each step of the Population II distance ladder.
 {\bf (top fork)} Illustration of the current \cchp~ distance ladder using the 5 trigonometric 
   parallaxes with \fgs resulting in a 2.9 per cent measure of \ho (Section \ref{currenterror}).
  Projected \cchp~ distance ladder error budget using $\sim$55 trigonometric
   parallaxes from \gaia resulting in a 2.4 per cent measure of \ho (Section \ref{gaia_rrl}).
 {\bf (bottom fork)} Projected \cchp~ distance ladder by-passing the RR~Lyrae and using
   the \hip parallaxes for local TRGB stars resulting
   in a 6.3 per cent measure of \ho(Section \ref{trgbonly}).
  Projected \cchp~ distance ladder using only the TRGB stars,
   but with improved \gaia parallaxes resulting in a 2.3 per cent measure of \ho (Section \ref{trgbonly}). 
  Despite minimizing the distance ladder and substantially reducing uncertainties, 
   the number of \sn zero point calibrators sets 
   the floor on the `local' measurement of \ho at 2.1 per cent.}
\end{figure*}  
\section{Discussion and Future Work} \label{sec:discussion}

The \cchp~ will provide a test of the Cepheid-based distance ladder with an equally 
 (or more) precise and systematically independent distance scale.   
We will produce an end-to-end calibration of the RR~Lyrae 
  distance scale in $F160W$.
Coupled with TRGB distances outside of the Local Group,
 the \cchp~ allows for a new and robust determination of the Hubble constant. 
In this section, we review our error budget (Section \ref{currenterror})
 and discuss future improvements for the calibration of the zero points for the 
 RRL PL relation (Section \ref{gaia_rrl}), TRGB (Section \ref{trgbonly}), and \sn (Section \ref{sn1a})
 and their impact on the measurement of \ho. 

\subsection{Summary \cchp~ Error Budget} \label{currenterror}

We now summarize the uncertainties for each of the steps in our current
 Population II distance ladder,
 as given in the black notation in the top fork of Figure \ref{chpflowchart}.
The \cchp~ has a geometric foundation calibrated by five Galactic RR Lyrae
 with \fgs parallaxes (Section \ref{sec:rrlyrae}),
 which calibrate the zero point of the RRL $F160W$ 
 PL relation to 1.7 per cent (first box in top fork of Figure \ref{chpflowchart}).
The megamaser distance to NGC\,4258 and RRL in six Local Group galaxies
 provide seven independent calibrations of the zero point for the TRGB (Section \ref{sec:trgbabs})
 to 0.9 per cent accuracy (second box in top fork of Figure \ref{chpflowchart}). 
The TRGB is then used to set the zero point for the local \sn sample,
 here in a set of nine host galaxies with a total of twelve individual \sn (Section \ref{sec:trgbsn1a}),
 resulting in a precision of 2.1 per cent (middle box of Figure \ref{chpflowchart}).
The \cchp~ can then tie into the current sample of over 200 (and growing) well-observed \sn 
 well into the Hubble flow (Section \ref{sec:hubbleflow})
 with a precision of 0.5 per cent (second box from right in Figure \ref{chpflowchart}).
Combining in quadrature each of the four systematic uncertainties in
 the Population~II route to the Hubble constant, 
 we conclude that a net uncertainty of 2.9 per cent is achievable with this program 
 (rightmost box for the top fork of Figure \ref{chpflowchart}).

An ancillary benefit of \cchp~ is the ability to compare this distance ladder to that from the Cepheids.
As indicated in Table \ref{tab:trgbtargets}, 
  five galaxies in our \sn sample and three in our Local Group sample
  also have distances via the Leavitt Law and, often,
  measured by multiple independent teams \citep[see summary in][]{freeman_2010, freedman_2012}. 
Thus, we can look for effects due to crowding, reddening, and metallicity in the Cepheid distances
 by comparing their results with those from this program 
 (see discussion in Section \ref{ssec:houncertainties}).
Moreover, we gain additional leverage on the calibration of the \sn zero point by 
 combining host galaxies with distances determined from both Population I and Population II distance indicators,
 though this is a process that requires extreme care as it combines measurements
 with very different systematic uncertainties.
Thus, we anticipate the greatest gain from the existence of two independent distance ladders
 is the ability to test individually stages of either distance ladder with a careful and systematic approach.
In the near future, however, there are significant opportunities to improve stages of both distance ladders
 (though we focus the following discussion to the Population II route described herein). 

\subsection{\emph{Gaia} Parallaxes for RR~Lyrae Variables} \label{gaia_rrl} 

As summarized in \citet[][and references therein]{clementini_2016},
 \gaia is expected to observe over one hundred thousand Galactic RRL, 
 a sample increase of several orders of magnitude compared to the 186 observed by \hip. 
According to \gaia's post-commissioning performance estimates \citep[see Table 1 in][]{debruijne_2014}, 
 all RRL with $\langle V \rangle$ \textless 12.1 mag 
 will have the parallax measured to better than $\sim$ 10 microarcsec, 
 whereas individual accuracies will range between 17 to 140 microarcsec for RRL in Galactic globular clusters 
 with typical $\langle V_{HB} \rangle$ between 14 and 18 mag. 
This implies that in a few years there will be more than an order of magnitude more RRL PL zero point calibrators. 
Moreover, owing to improvements in period coverage and metallicity spanned by this larger sample, 
 it may even be feasible to calibrate all terms in the period--luminosity--metallicity relation directly from the calibrators.

According to the current plan\footnote{The release plans are available here: \\ \url{http://www.cosmos.esa.int/web/gaia/release}}, 
 with the first \gaia release scheduled for summer 2016, 
 parallaxes with sub-milliarcsecond accuracy for the stars in common between the Tycho-2 Catalog and \gaia will be available soon. 
This initial catalog will be based on the Tycho-\gaia Astrometric Solution \citep[TGAS; see][]{michalik_2015} that will compare positions between the \gaia first release
 and the Tycho-2 catalog. 
Folding the \gaia trigonometric parallaxes into our Population II distance scale, however, requires additional observations.

In anticipation of the parallaxes being obtained by \gaia, 
  we are obtaining high quality $B$,$V$,$I$ optical light curves
  of the quality demonstrated for RR Gem in Figures \ref{fig:template}a and Figures \ref{fig:template}b
  for 55 Galactic RRL with the Three-hundred Millimeter Magellan Telescope 
 (TMMT; Monson et al. in prep).
A completed \hst Cycle 21 SNAP program to obtain $F160W$ magnitudes 
 includes 25 additional Galactic RRL calibrators \citep[PID 13472][]{hst_2013},
 for a total sample of 30 such stars when combined with the \cchp~ presented in Section \ref{ssec:plzp}.
These same 55 stars were also included in the CRRP \citep[][Scowcroft et al. in prep]{spitzer_2012}
 and have light curves in the MIR similar to that of RR Gem in Figure \ref{fig:template}d. 
When the trigonometric parallaxes become available,
 these additional Galactic calibrators can be seamlessly folded into 
 the framework described in Section \ref{sec:rrlyrae} to set the RRL foundation for the \cchp.

Beyond the gains for the Galactic calibrators, 
 \gaia will also provide distances to better than 1 per cent
  for about 77 Galactic globular clusters \citep[49 per cent of the known clusters and all clusters within 16 kpc;][]{cacciari_2013}.
Thus, \gaia will directly probe the RRL population within \ocen and other 
 clusters within $\sim$16 kpc of the Sun.
When coupled with chemical abundances derived from high resolution spectroscopy, 
 the larger sample of RRL within \gaia has the potential
 to resolve concerns over how metallicity affects the PL, both for the slope and zero point 
 (i.e., discussions in Section \ref{ssec:rrlmetals})
 that will further reduce the uncertainty associated with the use of RRL.

The order-of-magnitude increase in parallax calibrators and the
 more than an order-of-magnitude decrease in mean parallax error could drop the uncertainty
 on the RRL zero point from 1.7 per cent to 0.5 per cent in distance
 (0.079/$\sqrt{55}$ = 0.011~mag; indicated by red text in the leftmost box in upper fork of Figure \ref{chpflowchart};
  note we have adopted the zero point fitting uncertainty from Section \ref{ssec:plzp} as an upper limit). 
Propagating this change through to the measurement of \ho reduces the
 total predicted uncertainty on \ho from 2.9 to 2.4 per cent 
 (indicated by red text in the rightmost box in upper fork of Figure \ref{chpflowchart}).
 
\subsection{By-passing the RR~Lyrae Variables} \label{trgbonly} 

As previously discussed, the current systematic error on the absolute TRGB magnitude 
 is 0.12 mag or 6 per cent in distance \citep{rizzi_2007}
 and, thus, going directly to the TRGB at present is not an improvement
 from our current two-step process using the RRL 
 (black notations in the  forks of Figure \ref{chpflowchart}).
Moreover, using the current TRGB calibration results in a 6.3 per cent 
 measure of \ho that is not currently competitive with our current
 path ($\sim$3.0 per cent measure of \ho).
Ultimately, if we wish to calibrate the TRGB directly to better than 1 per cent,
 we must construct a distance ladder that relies only on the TRGB technique 
 from our local environment direct to the \sn calibrators.
  
\citet{tabur_2009} have combined \hip \citep{perryman_1997} and 
 photometry from the 2-Micron All Sky Survey \citep[2MASS][]{skrutskie_2006} to calibrate the 
 magnitude of the TRGB in the NIR ($K_s$). 
An optical all-sky catalog with comparable photometric precision does not yet exist.
Following the example of \citet{tabur_2009}, we are producing high quality optical photometry 
 for as many tip stars as are accessible from Las Campanas Observatory. 
More than 1000 of the \citet{tabur_2009} tip stars (determined from \hip) are accessible
 by the TMMT and we are pursuing $B$,$V$,$I$ photometry for these stars 
 (with multiple epochs to test for variable AGB contaminants at the tip).

The current \cchp~ TMMT sample contains over 1000 stars and the resulting TMMT $I$-band magnitudes 
 (part way through the survey) have mean photometric error of $\sigma_{I}\sim$0.04 mag
 (with removal of stars showing variability).
Unfortunately, the current \hip~ parallaxes have a mean $\sigma_{\pi}/\pi$ = 30 per cent
 and the average $M_{I}$ is only accurate to $\sigma_{M_{I}}\sim$ 0.9 mag or 45 per cent in distance.
Conservatively assuming $\sim$50 bona-fide tip stars, 
 this results in an uncertainty of 0.128 mag (5.9 per cent in distance),
 which offers no improvement on prior isochrone-based TRGB calibrations 
 (Section \ref{sec:trgbabs}). 

The median magnitude of our TRGB sample is $I$=5.34 (the average is $\langle I \rangle$ = 7.4), 
 which also falls well within the highest precision \gaia sample 
 \citep[$\sigma_{\pi}$ \textless ~7 micro-arcseconds for 3~ \textless~G~\textless ~12.1 mag; see][for end of mission values]{debruijne_2014}.
Assuming the fractional error of the trigonometric parallaxes ($\pi$) is reduced
 10-fold for our sources (a conservative projection), the mean $\sigma_{\pi}/\pi$ = 3 per cent.
Using our mean $I$ magnitude error ($\sigma_{I}=0.04$) 
 and a conservative estimate of 50 stars at the tip, 
 \gaia produces a 0.5 per cent calibration of the TRGB magnitude.
\gaia will sample many more tip stars than \hip~ and with corresponding optical photometry 
 can be added to our calibration sample \citep[e.g.,][]{mignard_2003,mignard_2005}. 
Moreover, the associated spectroscopic data products ($R=11500$) from \gaia for 
 stars brighter than $G\sim16$ will permit a high precision
 {\it empirical} calibration of the metallicity dependence 
 that is fully empirical and relies on a metallicities derived from a homogeneous analysis
 \citep[i.e., not using isochrones as in previous calibrations as by][]{madore_2009,mager_2008}.

Using projections for a \gaia calibration of the TRGB we estimate a 2.6 percent measure of \ho
 is feasible (rightmost box of the bottom fork of Figure \ref{chpflowchart}).
The \gaia branch of the bottom fork of Figure \ref{chpflowchart} represents a route
 to \ho where the floor set by the uncertainty on \sn zero point is reached.
Upon the completion of the work described here,
 the number of \sn calibrators (i.e., those near enough for independent distance determination)
 becomes the single limiting factor in the `local' measure of \ho.

\subsection{The \sn Calibrator Sample} \label{sn1a}

Even anticipating the significant gains for the distance ladder provided by the \gaia parallaxes, 
  the total \ho error budget is driven by the relatively small number of local \sn calibrators 
  (see blue points in Figure \ref{fig:hubbleflow}).
More importantly, this step affects not just our Population II route,
  but also those using Cepheids \citep[e.g., S$H_0$ES and CHP;][]{riess_2011, freedman_2012}.
As discussed by \citet{freeman_2010}, the small number of \sn calibrators 
 is a function of several factors: 
  (i) nature's rate of \sn production, 
  (ii) the ability of the astronomical community to detect transient objects, and 
  (iii) the properties of the host galaxy that make the \sn an ideal calibrator.
While we have no control over the rate of \sn, the \cchp~ Population II distance ladder
 now permits the utilization of \sn found in early-type and in edge-on late-type host galaxies, 
 that would previously be precluded from the calibrator sample owing to limitations of
 distance measurements with Cepheids.
Detectability of \sn can only be improved with the increased number and efficiency  
 of transient surveys, for example the dual-hemisphere ASAS-SN \citep[e.g.,][]{shappee_2014} 
  and the full-sky multi-epoch monitoring provided by the \gaia satellite 
  (with alerts predicted to total over 6000 transient objects over its five year lifetime), 
 that permit a vastly more complete sample of nearby, bright \sn to be discovered. 
 
These discoveries can only be capitalized upon for the distance ladder under specific circumstances,
  i.e., that the host galaxy is amenable to the non-trivial \sn follow-up described in Section \ref{sec:trgbsn1a}
  and within the volume probed by independent distance measures.
There are two factors setting volume: (i) collecting area (aperture) and (ii) resolution.
For \emph{HST}, this volume (set by practical concerns) 
 is $\sim$35 Mpc for Cepheids, 
    $\sim$20 Mpc for the optical TRGB
 and $\sim$1 Mpc for the optical/NIR RR~Lyrae --- all of which require large numbers of orbits 
 (\textgreater 10 orbits, with the non-variable TRGB being the most overall observationally efficient). 
With a 7-fold increase in collecting area and a 4-fold increase in resolution, the 
 \emph{James Webb Space Telescope} will greatly extend the volume within \sn host galaxies 
 have independently determined distances.
Proposed 30-meter class optical telescopes equipped with adaptive optics, such as the Giant Magellan Telescope (\emph{GMT}), the Thirty Meter Telescope (\emph{TMT}), and the 
 European Extremely Large Telescope (\emph{E-ELT})
 will be able to surpass \emph{HST} resolution in the NIR and
 will boast at minimum a 150-fold increase in collecting area over \emph{HST},
  though these gains will be modulated by atmospheric conditions. 
The combination of an increased observational effort from transient surveys,
 the new large aperture and high resolution facilities,
 and the ability to follow-up \sn from nearly any host galaxy using our Population II techniques,
 demonstrates that the technology and infrastructure is available to increase 
 the number of \sn calibrators.

The fiducial precision required to place high level constraints on 
 the CMB modeling from these `local' techniques is to measure \ho to 1 per cent
 \citep[see discussions in][]{planck_2013}.
Using the minimal Population II ladder schematically visualized 
 in the bottom fork Figure \ref{chpflowchart} with a \gaia calibrated TRGB
 and no other changes to the methodology,
 reaching 1 per cent in \ho would require a \sn zero point calibrated
 to 0.7 per cent and would require $\sim$100 \sn calibrators. 
With 50 \sn calibrators, the \sn zero point is calibrated to 1.5 per cent
 and \ho measured to 1.7 per cent;
 with 25 \sn calibrators, the zero point is calibrated to 0.9 per cent
 and \ho to 1.3 per cent.
Estimating the anticipated number of nearby \sn is severely limited by Poisson noise
 \citep[and the relatively short operational lifetimes of all-sky transient programs -- 1-2 years][]{asassn_2014_stats}
 but a `handful' (two to five) can be anticipated on an annual basis \citep{gal-yam_2013}\footnote{Updated daily and available at \url{http://www.rochesterastronomy.org/supernova.html}}. 
Thus, it may be feasible in the next two decades to reach a $\sim$1 per cent 
 measure of \ho from the Population II distance ladder.

\section{Final Remarks}\label{sec:summary}

The goal of the \cchp~ is to measure \ho using a distance ladder constructed
 from only Population II distance indicators.
Initially, the ladder will contain three distance tracers:
 the NIR RRL PL within 1 Mpc, the optical TRGB within $\sim$20 Mpc,
 and the \sn well into the smooth Hubble flow.
This path to \ho, however, requires calibrating the $F160W$ PL for RRL
 and providing a high precision calibration of the TRGB zero point. 
Using reasonable estimates for our error budget (i.e., upper limits on individual measurements), 
 this four step ladder (top fork in Figure \ref{chpflowchart}) 
 could initially provide a 2.9 per cent measure of \ho with 
 data available at the present time. 

While our initial \cchp~ measure of \ho is competitive with 
 other distance ladder measures (see compilation in Table \ref{tab:ho} and Figure \ref{fig:hotime}), 
 we critically assess how to push the Population II distance ladder to higher precision
 in the near future. 
Accounting for improvements to the calibration of RRL PL relation from \gaia (and associated
 observational campaigns), 
 the \ho precision attainable from the \cchp~ distance ladder could improve to 2.4 per cent.
Further improvement on the \ho precision occurs by by-passing the RRL
 and using Galactic TRGB calibrators to set the geometric foundation of the Population II scale.
With conservative estimates, a TRGB-only distance ladder can approach an \ho precision of 2.3 per cent,
 a value that is dominated by the number of independently calibrated \sn. 

Resolving the discrepancy between the traditional Cepheid based distance ladder 
 and CMB modeling remains critical. 
The \cchp~ distance ladder is completely independent of the Cepheid distance ladder
 and, as such, will provide a systematic check on each stage of the traditional Cepheid ladder.
The precision required of the distance ladder measure of \ho to provide a quantitative
 assessment of these differences is often set to 1 per cent.
Pushing the precision of \ho to this level is currently hampered by the number
 of \sn host galaxies with independent distances (Figure \ref{chpflowchart}).
With conservative estimates on \sn rates from all sky transient surveys 
 and the increase in the volume accessible by stellar population distance measures 
 permitted by next generation facilities, 
 it is possible to achieve such precision within the next few decades.

\acknowledgements
 We thank the anonymous referee for comments that have improved the manuscript.
We acknowledge numerous helpful conversations with and insights from Eric Persson.
RLB thanks Chris Burns for assistance in making Figure \ref{fig:hubbleflow}
 and for helpful discussions on the body of \sn measurements.
RLB, ISJ, and BFM also thank the organizers of the {\it New Era of the Cosmic Distance Scale} summer school at the
 University of Tokyo for an enlightening exploration of modern distance measurement techniques and the future of such programs.
G.C. wishes to thank Marcella Marconi for discussions on the theoretical background on the luminosity-metallicity and the period-luminosity-metallicity relations followed by RR Lyrae in optical and NIR/MIR passbands, respectively.
Support for program \#13691 was provided by NASA through a grant from the Space Telescope Science Institute, which is operated by the Association of Universities for Research in Astronomy, Inc., under NASA contract NASA 5-26555.
G.B. thanks the Japan Society for the Promotion of Science for a research grant (L15518).
G.C. acknowledges support by PRIN-INAF2014, Excalibur's: EXtragalactic distance scale CALIBration Using first-Rank Standard candles  (P.I. G. Clementini).
M.G.L. and I.S.J. were supported by the NRF grant funded by the Korea Government (No.~2012R1A4A1029713).
This work is based on observations made with the \emph{Spitzer} Space Telescope, which is operated by the Jet Propulsion Laboratory, California Institute of Technology under a contract with NASA. Support for this work was provided by NASA through an award issued by JPL/Caltech.
This research has made use of the NASA/IPAC Extragalactic Database (NED), which is operated by the Jet Propulsion Laboratory, California Institute of Technology, under contract with the National Aeronautics and Space Administration.

{\it Facilities:} \facility{HST (ACS, WFC3/IR)}, \facility{Magellan:Baade (FourStar)}, \facility{Spitzer (IRAC)}

\bibliographystyle{apj}
\bibliography{ms}

\end{document}